\documentclass[journal]{IEEEtran}

\ifCLASSINFOpdf
\usepackage[pdftex]{graphicx}
\DeclareGraphicsExtensions{.pdf,.jpeg,.png}
\else
\usepackage[dvips]{graphicx}
\fi
\ifCLASSOPTIONcompsoc
\usepackage[caption=false, font=normalsize, labelfont=sf, textfont=sf]{caption,subfig}
\else
\usepackage[caption=false, font=footnotesize]{subfig}
\fi
\usepackage{epstopdf}
\usepackage[cmex10]{amsmath}
\usepackage{amssymb}
\usepackage{wasysym}
\usepackage{array}
\usepackage{cite}
\usepackage{color}
\usepackage{url}
\usepackage{bm}
\usepackage{enumerate}
\usepackage{epsfig,latexsym}
\usepackage{flushend}
\usepackage{cite}
\usepackage{algorithm}
\usepackage{algorithmic}
\usepackage{diagbox} 
\usepackage{booktabs}
\usepackage{stfloats}
\usepackage{cases}

\makeatletter

\renewcommand{\maketag@@@}[1]{\hbox{\m@th\normalsize\normalfont#1}}%

\makeatother

\begin{document}
%
\title{Tri-Hybrid Beamforming for Radiation-Center Reconfigurable Antenna Array: Spectral Efficiency and Energy Efficiency}

\author{Yinchen~Li,~\IEEEmembership{Graduate Student Member,~IEEE},~Chenhao~Qi,~\IEEEmembership{Senior Member,~IEEE},\\
Shiwen~Mao,~\IEEEmembership{Fellow,~IEEE},~and~Octavia~A.~Dobre,~\IEEEmembership{Fellow,~IEEE}

\thanks{Part of this work has been accepted by the IEEE Global Communications Conference, Taipei, Taiwan, Dec. 2025~\cite{Li2025Tri}.}
\thanks{Yinchen~Li and Chenhao~Qi are with the School of Information Science and Engineering, Southeast University, Nanjing 210096, China (e-mail: \{liyinchen,qch\}@seu.edu.cn).}
\thanks{Shiwen Mao is with the Department of Electrical and Computer Engineering, Auburn University, Auburn, AL 36849-5201, USA (e-mail:	smao@ieee.org).}
\thanks{Octavia A. Dobre is with the Faculty of Engineering and Applied
	Science, Memorial University, St. John's, NL A1C 5S7, Canada (e-mail: odobre@mun.ca).}
}

\markboth{}{}

\maketitle



\begin{abstract}
	In this paper, we propose a tri-hybrid beamforming (THBF) architecture based on the radiation-center (RC) reconfigurable antenna array (RCRAA), including the digital beamforming, analog beamforming, and electromagnetic (EM) beamforming, where the EM beamformer design is modeled as RC selection.
	Aiming at spectral efficiency (SE) maximization subject to the hardware and power consumption constraints, we propose a tri-loop alternating optimization (TLAO) scheme for the THBF design, where the digital and analog beamformers are optimized based on the penalty dual decomposition in the inner and middle loops, and the RC selection is determined through the coordinate descent method in the outer loop. Aiming at energy-efficiency (EE) maximization, we develop a dual quadratic transform-based fractional programming (DQTFP) scheme, where the TLAO scheme is readily used for the THBF design. To reduce the computational complexity, we propose the Lagrange dual transform-based fractional programming (LDTFP) scheme, where each iteration has a closed-form solution.
	Simulation results demonstrate the great potential of the RCRAA in improving both SE and EE. Compared to the DQTFP scheme, the LDTFP scheme significantly reduces the computational complexity with only minor performance loss.
\end{abstract}


\section{Introduction}
The future wireless systems are expected to support abundant emerging applications, such as autonomous vehicles, wireless brain-computer interactions, and multisensory extended reality services~\cite{Saad2020Vision}. To meet the unprecedented communication demands for ultra-high data rate and ultra-low latency, millimeter wave (mmWave) communications have received widespread interest from both industry and academia~\cite{chowdhury20206g}. The short wavelength of mmWave signal facilitates the miniaturization of massive multiple-input multiple-output (MIMO) arrays, making it feasible to compensate for the high path loss and support parallel transmission of multiple data streams~\cite{Qi2021Acquisition}. However, the ever-increasing array size poses significant challenges in terms of hardware complexity and power consumption, leading to substantial difficulties for practical deployment. Although the extensively investigated hybrid beamforming (HBF) architecture can mitigate this challenge through reducing the number of RF chains, the issues of high cost and high power consumption remain fundamentally unresolved due to numerous RF components, such as phase shifters (PSs)~\cite{Wang2024Tutorial}. As a result, new architectures with low cost and power consumption need to be explored.

The design degrees-of-freedom (DoFs) plays a critical role in improving the performance of wireless systems~\cite{Sridharan2015Degrees}. Essentially, the enlargement in array size implies increased design DoFs, enabling fine-grained beam management and superior interference mitigation. Therefore, increasing design DoFs within fixed array dimensions is a potential cost-effective approach for next-generation wireless communications. An intuitive method is antenna selection (AS), which has been widely investigated in the past decade~\cite{Nguyen2017Joint},~\cite{Asaad2018Massive},~\cite{Asaad2024Joint}. By selecting a subset of antennas with favorable channel conditions, the system performance can be improved with reduced power consumption. However, the AS system requires the deployment of a large number of candidate antennas, resulting in prohibitively high hardware costs. Besides, the candidate antennas are arranged in a given array with fixed geometry, leading to limited design DoFs when using small-scale arrays.

To overcome these issues, reconfigurable arrays with movable antennas (MAs) or fluid antennas (FAs) have garnered much attention in recent years, where the positions or physical shapes of antennas can be altered to accommodate improved channel conditions~\cite{Zhu2024Movable},~\cite{wong2020fluid}. In contrast to AS, MA and FA technology can avoid the deployment of a large-scale candidate antenna array, and provide more DoFs owing to the flexible movement of antennas within a defined region. To verify these advantages, the field-response model of MA systems is developed, where a single MA can achieve comparable performance to the multi-antenna systems, providing inspiration for the design of cost-effective and energy-saving architectures~\cite{Zhu2024Modeling}. For systems with multiple receiving FAs, the antenna positions and beamforming design are jointly optimized, where the transmit power can be significantly reduced subject to the communication quality of service constraints~\cite{Qin2024Antenna}. 
Despite the great potential of MA and FA technologies to enhance performance with reduced power consumption and hardware complexity, the practical implementation faces some crucial difficulties. Specifically, the mechanical regulation of MAs and FAs introduces position update latency, rendering them incompatible with rapidly time-varying channels~\cite{Shen2020Beam}. In addition, the hardware manufacturing difficulties and long-term stability persist as unresolved problems~\cite{Yang2025Flexible}.

Recently, reconfigurable antennas have attracted considerable research interest~\cite{Castellanos2025embracing},~\cite{Shlezinger2021Dynamic},~\cite{Chen2025DBRAA}. By modifying the current distribution across the antenna surface, reconfigurable antennas enable programmable control of electromagnetic (EM) properties, thereby facilitating superior adaptation to the dynamic EM propagation environment. Different from MAs and FAs relying on physical restructuring to indirectly adjust radiation characteristics, reconfigurable antennas offer direct electrical control of EM properties, supporting rapid response and strong feasibility. Additionally, various types of EM property reconfiguration, such as polarization states~\cite{Zhu2014Design}, radiation patterns~\cite{Hossain2017Parasitic}, and frequency responses~\cite{Omari2023Design}, provide greater DoFs than MAs and FAs, presenting significant potential for performance enhancement. Among various implementations, reconfigurable pixel antennas (RPAs) exhibit notable advantages including rapid response, low cost, and high scalability~\cite{Rodrigo2014pixel}. Through dynamically altering the inter-pixel connections across the parasitic layer, the current distribution of the antenna surface can be modified, leading to the flexible control of radiation characteristics. To support the parallel transmission of multiple data streams in MIMO systems, a reconfigurable antenna array can be designed based on the RPA technology~\cite{Ying2024Harnessing}. To explore the performance gains, the analog precoding for single-user communications is investigated, where a radiation pattern reconfiguration method is developed to improve channel capacity~\cite{Li2020Analog}. For multiuser communications in Sub-6G band, two distinct RPA-based MIMO architectures are proposed, where a two-stage approach for beamforming and antenna selection is designed~\cite{chen2025remaa}. For multiuser communications in mmWave band, it is significant to investigate the state-of-the-art tri-hybrid beamforming (THBF) architecture, by integrating reconfigurable antenna technology with the popular HBF architecture~\cite{Castellanos2025embracing}. However, the joint design of the digital beamforming, analog beamforming, and EM beamforming through antenna reconfiguration is still in the embryonic phase. In particular, the modeling of EM beamforming remains an open issue.



Both spectral efficiency (SE) and energy efficiency (EE) are critical performance metrics for communication systems. In particular, due to the increasing demand for carbon neutrality and sustainability, EE has become a critical consideration in the design of future wireless systems~\cite{Zhong2024Toward}. 
In this paper, we propose a THBF architecture based on the radiation-center (RC) reconfigurable antenna array (RCRAA), where we consider the THBF design for SE maximization or EE maximization. The main contributions are summarized as follows, where the first point is included in our conference paper~\cite{Li2025Tri}.

\begin{itemize}
	\item[1)] We propose the RCRAA-based THBF architecture, where the EM beamformer design is modeled as an RC selection problem. Then, we jointly design the digital, analog, and EM beamformers, aiming at SE maximization subject to the hardware and transmit power constraints. To solve this non-convex problem, we propose a tri-loop alternating optimization (TLAO) framework, where the analog and digital beamformers are optimized in the inner and middle loops, and the RC selection is determined through the coordinate descent (COD) method in the outer loop.
	\item[2)] We consider the THBF design aiming at EE maximization subject to the hardware and transmit power constraints, where the TLAO framework is readily used to solve it. We employ a general power consumption model to effectively compare and analysis the EE performance for different architectures.
	\item[3)] In the inner and middle loops of the TLAO framework, we propose an HBF scheme with fixed radiation centers (RCs) based on the penalty dual decomposition (PDD-HBF). Specifically, for SE maximization, we equivalently transform the original problem into the weighted minimum mean squared error (WMMSE) problem. For EE maximization, we develop a dual quadratic transform-based fractional programming (DQTFP) scheme to tackle the nested fractional form.
	\item[4)] To reduce the computational complexity of the DQTFP scheme, we propose a Lagrange dual transform-based fractional programming (LDTFP) scheme, where we use the epigraph problem to extract the inner fractional form from the logarithmic function. Different from the DQTFP scheme, each iteration of the LDTFP scheme has a closed-form solution, thereby resulting in low complexity.
\end{itemize}

The rest of this paper is organized as follows. The modeling of the RCRAA and the multiuser mmWave communication system is introduced in Section~\ref{section:System Model}. The THBF designs for SE and EE maximization are provided in Section~\ref{section:THB SR} and Section~\ref{section:THB EE}, respectively. The simulation results are presented in Section~\ref{section:simulation}. Finally, we will conclude this paper in Section~\ref{section:conclusion}.

\textit{\textbf{Notations:}} Symbols for vectors (lower case) and matrices (upper case) are in boldface. $(\cdot)^{\ast}$, $(\cdot)^{\rm T}$, and $(\cdot)^{\rm H}$ denote the conjugate, transpose, and conjugate transpose, respectively. $[\boldsymbol{a}]_{n}$ and $\|\boldsymbol{a}\|_{2}$ denote the $n$th entry and the $l_{2}$-norm of the vector $\boldsymbol{a}$, respectively. $[\boldsymbol{A}]_{m,n}$, $[\boldsymbol{A}]_{m,:}$, $[\boldsymbol{A}]_{:,n}$, and $\|\boldsymbol{A}\|_{\rm F}$ denote the entry on the $m$th row and $n$th column, the $m$th row, the $n$th column, and the Frobenius norm of the matrix $\boldsymbol{A}$, respectively. $j$, $\boldsymbol{1}_{N}$, and $\boldsymbol{I}_{N}$ denote the square root of $-1$, the $N\times 1$ identity vector, and the $N\times N$ identity matrix, respectively. $\mathbb{Z}$, $\mathbb{C}$, and $\mathbb{R}$ denote the set of integers, complex numbers, and real numbers, respectively. $\mathrm{blkdiag}(\cdot)$, $\mathrm{vec}(\cdot)$, $\mathrm{Re}(\cdot)$, $\angle(\cdot)$, and $\mathcal{CN}(\cdot)$ denote the block diagonalization, vectorization, real part, argument, and complex Gaussian distribution, respectively.

\section{System Model}\label{section:System Model}
\begin{figure}[!t]
	\begin{center}
		\includegraphics[width=85mm]{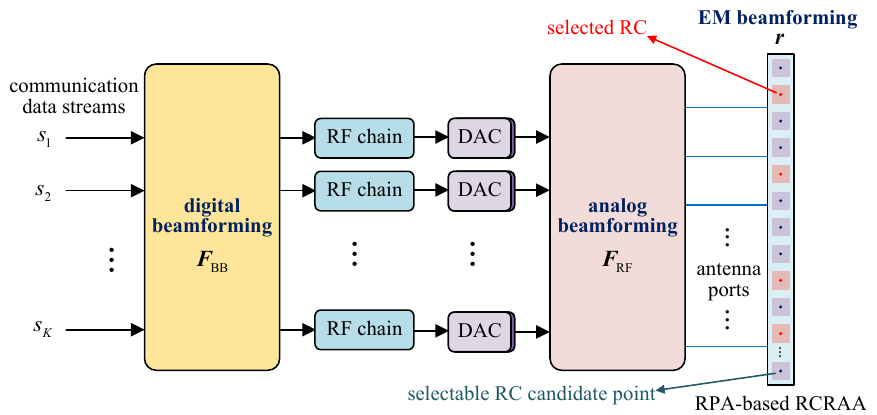}
	\end{center}
	\caption{Illustration of the RCRAA-based THBF architecture for multiuser mmWave communications.}\label{fig:system model}
\end{figure}

We consider a multiuser mmWave downlink communication system. As illustrated in Fig.~\ref{fig:system model}, the BS is equipped with an RCRAA-based THBF architecture, including the digital, analog, and EM beamforming. In the following, the implementation and modeling of the RCRAA will be presented in Section~\ref{subsection:RCRAA}. The signal propagation model and the performance metrics will be presented in Section~\ref{subsection:signal} and Section~\ref{subsection:performance metrics}, respectively.

\subsection{Implementation and Modeling of RCRAA}\label{subsection:RCRAA}
For multipath wireless channels, the receive signal strength depends on the interference among various multipath components, leading to spatial-varying channel gain. For multiuser mmWave communications, stronger signal strength usually indicates better channel condition and system performance. Besides, both the spatial distribution of users and the array geometry affect the capability to mitigate multiuser interference (MUI)~\cite{Chen2024Near}. To enlarge design DoFs for multiuser mmWave communications, we propose the RCRAA. Benefiting from the flexible RC selection, the RCRAA can adapt to favorable EM propagation environment to improve the system performance.

\begin{figure}[t]
	\begin{center}
		\includegraphics[width=80mm]{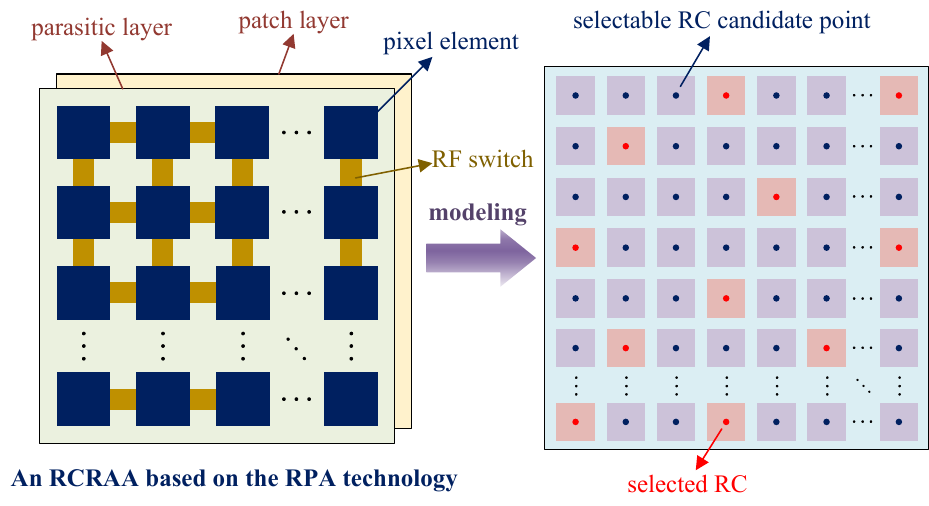}
	\end{center}
	\caption{Modeling of the RPA-based RCRAA.}\label{fig:RCRAA}
\end{figure}

The hardware implementation of the RCRAA includes MAs~\cite{Zhu2024MA}, FAs~\cite{wong2020fluid}, holographic metasurfaces~\cite{Huang2020Holographic}, and RPA technology~\cite{Zhang2025pixel}. 
Due to the superiority of RPAs in flexible and practical implementations, we provide an example of RPA-based RCRAA, which is implemented by an RPA array and composed of a patch layer and a parasitic layer~\cite{Ying2024Harnessing}. Specifically, the patch layer is connected to the RF network and equipped with multiple antenna ports, coupling the EM energy into the parasitic layer. The parasitic layer contains a large number of subwavelength-sized pixel elements interconnected with tunable RF switches, such as PIN diodes or vanadium dioxide components, radiating the EM energy into space~\cite{Soltani2018Design}.
Through modifying the inter-pixel connections, the radiation characteristics of the RCRAA can be altered, thereby adapting to favorable channel conditions~\cite{Rodrigo2014pixel}. Then multiple selectable RC candidate points can be equivalently discretized for the RCRAA~\cite{Zhang2025pixel}, as illustrated in Fig.~\ref{fig:RCRAA}. As a result, we model the EM beamformer design as selection of RCs from the RC candidate points.

\textbf{Remark 1:} In contrast to traditional AS technology based on a large number of fixed-position antennas (FPAs), the RCRAA enables finer-grained control on radiation characteristics, by selecting RCs from RC candidate points with subwavelength spacing. The implementation of the RCRAA does not require the deployment of numerous antenna ports, PSs, and power amplifiers (PAs), indicating that the RCRAA can increase the design DoFs with low power consumption and hardware complexity.

\subsection{Signal Propagation Model}\label{subsection:signal}
We denote the number of downlink single-antenna users by $K$. For the mmWave BS, we denote the numbers of RF chains, antenna ports, and selectable RC candidate points as $N_{\rm RF}$, $N_{\rm T}$, and $N_{\rm EM}$, respectively, satisfying $K=N_{\rm RF}\leq N_{\rm T}\leq N_{\rm EM}$. For simplicity, the selectable RC candidate points are arranged in a uniform linear array (ULA) with an interval of $d_{\rm p}$. Although this work focuses on 2D scenario, the proposed scheme can be readily applied to 3D scenario by extending the ULA to a planar array.

The RC selection vector is denoted by $\boldsymbol{r}\in\mathbb{Z}^{N_{\rm EM}}$ and satisfies
\begin{equation}\label{r_con:binary}
	[\boldsymbol{r}]_{n}\in\{0,1\},
\end{equation}
for $n=1,2,\cdots,N_{\rm EM}$, where $[\boldsymbol{r}]_{n}=1$ or $[\boldsymbol{r}]_{n}=0$ indicates that the $n$th selectable RC candidate point is selected or not, respectively. Due to the hardware connection limitations between antenna ports and pixel elements, only $N_{\rm T}$ RCs can be selected from totally $N_{\rm EM}$ selectable RC candidate points. This constraint can be expressed as
\begin{equation}\label{r_con:port}
	\boldsymbol{1}_{N_{\rm EM}}^{\rm T}\boldsymbol{r}=N_{\rm T}.
\end{equation}
The carrier wavelength of the system is denoted by $\lambda$. To avoid mutual coupling effect, the spacing between adjacent RCs should be no less than $\lambda/2$, which is equivalent to the minimum index spacing of $D_{\rm min}\triangleq\left\lceil\lambda/(2d_{\rm p})\right\rceil$. This constraint indicates that at most one RC can be chosen from any $D_{\rm min}$ adjacent selectable RC candidate points. We define a binary vector $\boldsymbol{b}_{n}$ as
\begin{align}
	\left[\boldsymbol{b}_{n}\right]_{i}\triangleq\left\{\begin{aligned}
		1&,~~n\leq i\leq n+D_{\rm min}-1,\\
		0&,~~{\rm others},
	\end{aligned}\right.
\end{align}
for $n=1,2,\cdots,N_{\rm EM}-D_{\rm min}+1$, and further define $\boldsymbol{B}\triangleq[\boldsymbol{b}_{1},\boldsymbol{b}_{2},\cdots,\boldsymbol{b}_{\widetilde{N}_{\rm EM}}]$, where we set $\widetilde{N}_{\rm EM}\triangleq N_{\rm EM}-D_{\rm min}+1$. Then, $\boldsymbol{r}$ should also satisfy
\begin{equation}\label{r_con:mutual coupling}
	\boldsymbol{B}^{\rm T}\boldsymbol{r}\leq \boldsymbol{1}_{\widetilde{N}_{\rm EM}}.
\end{equation}
Therefore, the feasible set of $\boldsymbol{r}$ can be defined as
\begin{equation}\label{A_r:RC feasible set}
	\mathcal{A}_{r}\triangleq\left\lbrace\boldsymbol{r}\in\mathbb{Z}^{N_{\rm EM}}\middle|~\eqref{r_con:binary},~\eqref{r_con:port},~{\rm and}~\eqref{r_con:mutual coupling}\right\rbrace.
\end{equation}

Considering all the selectable RC candidate points in the RCRAA, we denote the dimension-extended mmWave channel between the BS and users as $\bar{\boldsymbol{H}}\triangleq[\bar{\boldsymbol{h}}_{1},\bar{\boldsymbol{h}}_{2},\cdots,\bar{\boldsymbol{h}}_{K}]$. Based on the Salah-Valenzuala model, we can express $\bar{\boldsymbol{h}}_{k}\in\mathbb{C}^{N_{\rm EM}}$ as
\begin{equation}\label{hkbar:dimension-extended channel}
	\bar{\boldsymbol{h}}_{k}\triangleq\sqrt{\frac{N_{\rm EM}}{L_{k}}}\sum_{l=1}^{L_{k}}\beta_{k}^{(l)}\boldsymbol{\alpha}\left(N_{\rm EM},\theta_{k}^{(l)},d_{\rm p}\right),
\end{equation}
where $L_{k}$ denotes the numbers of channel paths. $\beta_{k}^{(l)}$ and $\theta_{k}^{(l)}$ denote the channel gain and the angle-of-departure of the $l$th path, respectively. $\boldsymbol{\alpha}(N,\theta,d)$ denotes the normalized array response vector expressed as
\begin{equation}
	\boldsymbol{\alpha}\left(N,\theta,d\right)=\frac{1}{\sqrt{N}}[1,e^{j\frac{2\pi}{\lambda}d{\rm sin}\theta},\cdots,e^{j\frac{2\pi}{\lambda}(N-1)d{\rm sin}\theta}]^{\rm T}.
\end{equation}

In fact, the practical channel between the BS and users, denoted by $\boldsymbol{H}(\boldsymbol{r})\triangleq[\boldsymbol{h}_{1}(\boldsymbol{r}),\boldsymbol{h}_{2}(\boldsymbol{r}),\cdots,\boldsymbol{h}_{K}(\boldsymbol{r})]\in\mathbb{C}^{N_{\rm T}\times K}$, is determined by the RC selection. We define the index set of the selected RCs as $\mathcal{X}\triangleq\{x_{1},x_{2},\cdots x_{N_{\rm T}}\}$. The mapping from $\mathcal{X}$ to $\boldsymbol{r}$ can be expressed as
\begin{align}\label{mapping:X to r}
	\left[\boldsymbol{r}\right]_{n}=\left\{\begin{aligned}
			1&,~~n\in\mathcal{X},\\
			0&,~~{\rm others}.
		\end{aligned}\right.
\end{align}
Then, we define $\boldsymbol{T}\in\mathbb{Z}^{N_{\rm T}\times N_{\rm EM}}$ satisfying
\begin{subequations}
	\begin{align}
		&\left[\boldsymbol{T}\right]_{m,n}=\left\{\begin{aligned}
			1&,~~n\in\mathcal{X},\\
			0&,~~{\rm others},
		\end{aligned}\right.\\
		&\sum_{n=1}^{N_{\rm EM}}\left[\boldsymbol{T}\right]_{m,n}=1,~~\sum_{m=1}^{N_{\rm T}}\left[\boldsymbol{T}\right]_{m,n}\leq1,
	\end{align}	
\end{subequations}
for $m=1,2,\cdots,N_{\rm T}$, and $n=1,2,\cdots,N_{\rm EM}$. As a result, $\boldsymbol{h}_{k}(\boldsymbol{r})$ can be expressed as a linear function of $\boldsymbol{r}$, i.e.,
\begin{equation}\label{hkr:practical channel}
	\boldsymbol{h}_{k}(\boldsymbol{r})=\boldsymbol{T}\bar{\boldsymbol{h}}_{k}.
\end{equation}

The baseband communication data is denoted by $\boldsymbol{s}\triangleq[s_{1},s_{2},\cdots,s_{K}]^{\rm T}$, satisfying $\mathbb{E}\{\boldsymbol{s}\boldsymbol{s}^{\rm H}\}=\boldsymbol{I}_{K}$. For $k=1,2,\cdots,K$, the received signal of the $k$th user can be expressed as
\begin{equation}\label{y_k:received signal}
	y_{k}=\boldsymbol{h}_{k}^{\rm H}(\boldsymbol{r})\boldsymbol{F}_{\rm RF}\boldsymbol{F}_{\rm BB}\boldsymbol{s}+n_{k},
\end{equation}
where $\boldsymbol{F}_{\rm RF}\in\mathbb{C}^{N_{\rm T}\times K}$, $\boldsymbol{F}_{\rm BB}\in\mathbb{C}^{K\times K}$, and $n_{k}\in\mathcal{CN}(0,\sigma_{n}^{2})$ denote the analog beamformer, the digital beamformer, and additive white Gaussian noise, respectively.

Since the analog beamforming is implemented with PSs, each non-zero entry of $\boldsymbol{F}_{\rm RF}$ satisfies the constant modulus constraint. Considering both fully-connected hybrid beamformer (FC-HBF) and partially-connected hybrid beamformer (PC-HBF), we define $\mathcal{F}_{\rm FC}$ and $\mathcal{F}_{\rm PC}$ as the feasible sets of $\boldsymbol{F}_{\rm RF}$, respectively. We can express $\mathcal{F}_{\rm FC}$ and $\mathcal{F}_{\rm PC}$ as \eqref{FFC:feasible set} and \eqref{FPC:feasible set} shown at the top of this page, respectively, where we define~$N_{\rm s}\triangleq \left\lfloor N_{\rm T}/K \right\rfloor$.

\setcounter{equation}{11}
\begin{figure*}[!t]
	\begin{equation}\label{FFC:feasible set}
		\mathcal{F}_{\rm FC}\triangleq\left\lbrace\boldsymbol{F}_{\rm RF}\in\mathbb{C}^{N_{\rm T}\times K}\middle|~\big|\left[\boldsymbol{
			F}_{\rm RF}\right]_{m,n}\big|=1\right\rbrace.
	\end{equation}
\end{figure*}
\setcounter{equation}{13}

\setcounter{equation}{12}
\begin{figure*}
	\vspace*{-10pt}
	\begin{equation}\label{FPC:feasible set}
		\mathcal{F}_{\rm PC}\triangleq\left\lbrace\boldsymbol{F}_{\rm RF}\in\mathbb{C}^{N_{\rm T}\times K}\middle|\boldsymbol{F}_{\rm RF}={\rm blkdiag}\{\boldsymbol{f}_{1}^{\rm RF},\boldsymbol{f}_{2}^{\rm RF},\cdots,\boldsymbol{f}_{K}^{\rm RF}\},~\boldsymbol{f}_{k}^{\rm RF}\in\mathbb{C}^{N_{\rm s}},~\big|\left[\boldsymbol{f}_{k}^{\rm RF}\right]_{m}\big|=1\right\rbrace.
	\end{equation}
	\rule[5pt]{18.1cm}{0.05em}
	\vspace*{-20pt}
\end{figure*}
\setcounter{equation}{14}

\subsection{Performance Metrics}\label{subsection:performance metrics}
We use SE and EE as the performance metrics for the multiuser mmWave system. Based on \eqref{y_k:received signal}, we can express the signal-to-interference-plus-noise ratio (SINR) of the $k$th user as
\setcounter{equation}{13}
\begin{equation}\label{gamma_k:SINR}
	\gamma_{k}=\frac{\left|\boldsymbol{h}_{k}^{\rm H}(\boldsymbol{r})\boldsymbol{F}_{\rm RF}\left[\boldsymbol{F}_{\rm BB}\right]_{:,k}\right|^{2}}{\sum_{i=1,i\neq k}^{K}\left|\boldsymbol{h}_{k}^{\rm H}(\boldsymbol{r})\boldsymbol{F}_{\rm RF}\left[\boldsymbol{F}_{\rm BB}\right]_{:,i}\right|^{2}+\sigma_{n}
	^{2}}.
\end{equation}
Then, the SE of the $K$ users can be calculated as
\begin{equation}\label{Rsum:sum-rate}
	R_{\rm SE}=\sum_{k=1}^{K}{\rm log}_{2}\left(1+\gamma_{k}\right).
\end{equation}

The transmit power of the BS, denoted by $P_{\rm T}$, can be expressed as
\begin{equation}\label{PT:transmit power}
	P_{\rm T}=\mathbb{E}\big\{{\rm tr}(\boldsymbol{F}_{\rm RF}\boldsymbol{F}_{\rm BB}\boldsymbol{s}\boldsymbol{s}^{\rm H}\boldsymbol{F}_{\rm BB}^{\rm H}\boldsymbol{F}_{\rm RF}^{\rm H})\big\}=\|\boldsymbol{F}_{\rm RF}\boldsymbol{F}_{\rm BB}\|_{\rm F}^{2}.
\end{equation}
We denote the total power consumption as $P_{\rm total}$, which is jointly determined by transmit power and RF component power. We can express $P_{\rm total}$ as
\begin{equation}
	P_{\rm total}=\frac{P_{\rm T}}{\eta_{\rm PA}}+P_{\rm LO}+N_{\rm RF}\left(P_{\rm RF}+2P_{\rm DAC}\right)+N_{\rm PS}P_{\rm PS},
\end{equation}
where $P_{\rm LO}$, $P_{\rm RF}$, $P_{\rm DAC}$, and $P_{\rm PS}$ represent the power consumption of the local oscillator, RF chain, digital-to-analog converter (DAC), and PS, respectively~\cite{Castellanos2025embracing}. $\eta_{\rm PA}$ and $N_{\rm PS}$ denote the PA efficiency and the number of PSs, respectively. For the FC-HBF and PC-HBF, we have $N_{\rm PS}=N_{\rm T}N_{\rm RF}$ and $N_{\rm PS}=N_{\rm T}$, respectively. Then, the EE can be defined as
\begin{equation}\label{etaEE:definition}
	\eta_{\rm EE}\triangleq\frac{R_{\rm SE}}{P_{\rm total}}.
\end{equation}

In the following sections, we will focus on the THBF design aiming at SE or EE maximization based on the defined performance metrics.

\section{THBF Design for SE Maximization}\label{section:THB SR}
\subsection{Problem Formulation}
To maximize the SE subject to the hardware and transmit power constraints, we formulate the optimization problem as
\begin{subequations}\label{max:sum-rate}
	\begin{align}
		\underset{\boldsymbol{r},\boldsymbol{F}_{\rm RF},\boldsymbol{F}_{\rm BB}}{\max}~~&\sum_{k=1}^{K}{\rm log}_{2}\left(1+\gamma_{k}\right)\label{max:sum-rate:objective}\\
		\mathrm{s.t.}~~~~~~&\boldsymbol{r}\in\mathcal{A}_{r},\label{max:sum-rate:RC constraint}\\
		&\boldsymbol{F}_{\rm RF}\in\mathcal{F}_{\rm FC}~{\rm or}~\mathcal{F}_{\rm PC},\label{max:sum-rate:FRF constraint}\\
		&\|\boldsymbol{F}_{\rm RF}\boldsymbol{F}_{\rm BB}\|_{\rm F}^{2}\leq P_{\rm max},\label{max:sum-rate:PT constraint}
	\end{align}
\end{subequations}
where~\eqref{max:sum-rate:RC constraint}, \eqref{max:sum-rate:FRF constraint}, and \eqref{max:sum-rate:PT constraint} denote the constraints of the RC selection, the analog beamformer, and transmit power, respectively. $P_{\rm max}$ denotes the maximum transmit power of the BS. The expressions of $\mathcal{A}_{r}$, $\mathcal{F}_{\rm FC}$, $\mathcal{F}_{\rm PC}$, and $\gamma_{k}$, for $k=1,2,\cdots,K$, are presented in~\eqref{A_r:RC feasible set}, \eqref{FFC:feasible set}, \eqref{FPC:feasible set}, and \eqref{gamma_k:SINR}, respectively. Note that \eqref{max:sum-rate} is difficult to solve owing to the non-convex objective function, the discrete constraint in~\eqref{max:sum-rate:RC constraint}, the constant-modulus constraint in~\eqref{max:sum-rate:FRF constraint}, and the coupling of $\boldsymbol{r}$, $\boldsymbol{F}_{\rm RF}$, and $\boldsymbol{F}_{\rm BB}$. We will decompose this problem into several tractable subproblems and propose a TLAO framework for solving.

\subsection{WMMSE-Based PDD-HBF Scheme}\label{subsection:PDD-HBF based on WMMSE}
We first focus on the design of $\boldsymbol{F}_{\rm RF}$ and $\boldsymbol{F}_{\rm BB}$ with fixed $\boldsymbol{r}$. By introducing an auxiliary matrix $\boldsymbol{F}\triangleq\boldsymbol{F}_{\rm RF}\boldsymbol{F}_{\rm BB}$, we can rewrite~\eqref{gamma_k:SINR},~\eqref{max:sum-rate:objective}, and~\eqref{max:sum-rate:PT constraint} as
\begin{equation}\label{gamma_k widetilde}
	\widetilde{\gamma}_{k}=\frac{\left|\boldsymbol{h}_{k}^{\rm H}(\boldsymbol{r})\boldsymbol{f}_{k}\right|^{2}}{\sum_{i=1,i\neq k}^{K}\left|\boldsymbol{h}_{k}^{\rm H}(\boldsymbol{r})\boldsymbol{f}_{i}\right|^{2}+\sigma_{n}^{2}},~~k=1,2,\cdots,K,
\end{equation}
\begin{equation}
	\widetilde{R}_{\rm sum}=\sum_{k=1}^{K}{\rm log}_{2}\left(1+\widetilde{\gamma}_{k}\right),
\end{equation}
\begin{equation}\label{AUX:PT constraint}
	\|\boldsymbol{F}\|_{\rm F}^{2}\leq P_{\rm max},
\end{equation}
respectively, where we define $\boldsymbol{f}_{k}\triangleq\left[\boldsymbol{F}\right]_{:,k}$. Then the HBF optimization problem with fixed $\boldsymbol{r}$ can be reformulated as
\begin{subequations}\label{max:sum-rate AUX}
	\begin{align}
		\underset{\boldsymbol{F},\boldsymbol{F}_{\rm RF},\boldsymbol{F}_{\rm BB}}{\max}~~&\widetilde{R}_{\rm sum}\label{max:sum-rate AUX:objection}\\
		\mathrm{s.t.}~~~~~~&\boldsymbol{F}=\boldsymbol{F}_{\rm RF}\boldsymbol{F}_{\rm BB}\label{AUX:coupling equality constraint}\\
		&\eqref{max:sum-rate:FRF constraint}~\mathrm{and}~\eqref{AUX:PT constraint}.
	\end{align}
\end{subequations}

Due to the equivalence between the SE maximization and the WMMSE problems~\cite{zhao2023rethinking}, we can convert~\eqref{max:sum-rate AUX} to
\begin{subequations}\label{min:WMMSE}
	\begin{align}
		\underset{\boldsymbol{F},\boldsymbol{F}_{\rm RF},\boldsymbol{F}_{\rm BB},\{w_{k},u_{k}\}_{k=1}^{K}}{\min}~~&\sum_{k=1}^{K}\left(w_{k}e_{k}-{\rm log}_{2}w_{k}\right),\label{min:WMMSE:objective}\\
		\mathrm{s.t.}~~~~~~~~~~~~&\eqref{AUX:coupling equality constraint},~\eqref{max:sum-rate:FRF constraint},\mathrm{and}~\eqref{AUX:PT constraint},
	\end{align}
\end{subequations}
where $w_{k}>0$ and $u_{k}$ are auxiliary variables, for $k=1,2,\cdots,K$. We can express $e_{k}$ as
\begin{align}
	e_{k}=&\sum_{i=1,i\neq k}^{K}\left|u_{k}\boldsymbol{h}_{k}^{\rm H}(\boldsymbol{r})\boldsymbol{f}_{i}\right|^{2}+\left|u_{k}\right|^{2}\sigma_{n}^{2}\nonumber\\
	&+\left|1-u_{k}^{\ast}\boldsymbol{f}_{k}^{\rm H}\boldsymbol{h}_{k}(\boldsymbol{r})\right|^{2}.
\end{align}

To tackle the coupling equality constraint~\eqref{AUX:coupling equality constraint} based on the penalty dual decomposition (PDD) method, we incorporate~\eqref{AUX:coupling equality constraint} as a penalty term into~\eqref{min:WMMSE:objective}. We can rewrite~\eqref{min:WMMSE:objective} as~\cite{Shi2020Penalty}
\begin{align}\label{L1:WMMSE+PDD}
	&L_{1}\left(\boldsymbol{F},\boldsymbol{F}_{\rm RF},\boldsymbol{F}_{\rm BB},\boldsymbol{w},\boldsymbol{u}\right)\nonumber\\
	\triangleq&\sum_{k=1}^{K}\left(w_{k}e_{k}-\mathrm{log}_{2}w_{k}\right)+\frac{1}{2\mu_{1}}\|\boldsymbol{F}-\boldsymbol{F}_{\rm RF}\boldsymbol{F}_{\rm BB}\|_{\rm F}^{2},
\end{align}
where $\mu_{1}>0$ is the penalty coefficient. We define $\boldsymbol{w}\triangleq[w_{1},w_{2},\cdots,w_{K}]^{\rm T}$ and $\boldsymbol{u}\triangleq[u_{1},u_{2},\cdots,u_{K}]^{\rm T}$. Thus, we can transform~\eqref{min:WMMSE} into
\begin{subequations}\label{min:WMMSE PDD}
	\begin{align}
		\underset{\boldsymbol{F},\boldsymbol{F}_{\rm RF},\boldsymbol{F}_{\rm BB},\boldsymbol{w},\boldsymbol{u}}{\min}~~&L_{1}\left(\boldsymbol{F},\boldsymbol{F}_{\rm RF},\boldsymbol{F}_{\rm BB},\boldsymbol{w},\boldsymbol{u}\right)\label{min:WMMSE PDD:objective}\\
		\mathrm{s.t.}~~~~~~~~&\eqref{max:sum-rate:FRF constraint}~\mathrm{and}~\eqref{AUX:PT constraint}.
	\end{align}
\end{subequations}
Then, we adopt the PDD-HBF scheme to solve~\eqref{min:WMMSE PDD} in a dual-loop framework. Specifically, we update $\mu_{1}$ in the outer loop, and optimize $\boldsymbol{u}$, $\boldsymbol{w}$, $\boldsymbol{F}$, $\boldsymbol{F}_{\rm RF}$, and $\boldsymbol{F}_{\rm BB}$ alternately in the inner loop. The detailed procedures are described as follows.

\emph{1) Initialization:} Based on the maximum ratio transmission theory~\cite{Lo1999Maximum}, we can initialize $\boldsymbol{F}$ as
\begin{equation}\label{min:WMMSE:initialize F}
	\boldsymbol{F}=\frac{\sqrt{P_{\rm max}}}{\|\boldsymbol{H}(\boldsymbol{r})\|_{\rm F}}\boldsymbol{H}(\boldsymbol{r}).
\end{equation}
For the FC-HBF and PC-HBF, we can respectively initialize $\boldsymbol{F}_{\rm RF}$ as
\begin{equation}\label{min:WMMSE:initialize FRF FC}
	\boldsymbol{F}_{\rm RF}=\exp\left(j\angle\left(\boldsymbol{H}(\boldsymbol{r})\right)\right),
\end{equation}
\begin{equation}\label{min:WMMSE:initialize FRF PC}
	\boldsymbol{f}_{k}^{\rm RF}=\exp\left(j\angle\left(\left[\boldsymbol{h}_{k}(\boldsymbol{r})\right]_{(k-1)N_{\rm s}+1:kN_{\rm s}}\right)\right),
\end{equation}
where $\boldsymbol{f}_{k}^{\rm RF}$ is the $k$th diagonal block of $\boldsymbol{F}_{\rm RF}$ as presented in~\eqref{FPC:feasible set}, for $k=1,2,\cdots,K$. We initialize $\boldsymbol{F}_{\rm BB}$ as a diagonal matrix expressed as
\begin{equation}\label{min:WMMSE:initialize FBB}
	\boldsymbol{F}_{\rm BB}=\frac{\sqrt{P_{\rm max}}}{\|\boldsymbol{F}_{\rm RF}\boldsymbol{I}_{K}\|_{\rm F}}\boldsymbol{I}_{K}.
\end{equation}

\emph{2) Optimization of $\boldsymbol{u}$:} Since the optimization of each component in $\boldsymbol{u}$ is independent, the subproblem to optimize $u_{k}$ can be expressed as
\begin{align}
	\underset{u_{k}}{\min}~~e_{k}\triangleq&\left|u_{k}\right|^{2}\big(\sum_{i=1}^{K}\left|\boldsymbol{h}_{k}^{\rm H}(\boldsymbol{r})\boldsymbol{f}_{i}\right|^{2}+\sigma_{n}^{2}\big)+1\nonumber\\
	&-2\mathrm{Re}\{u_{k}\boldsymbol{h}_{k}^{\rm H}(\boldsymbol{r})\boldsymbol{f}_{k}\},
\end{align}
where $e_{k}$ is a convex function of $u_{k}$, for $k=1,2,\cdots,K$. By setting $\partial e_{k}/\partial u_{k}^{\ast}=0$, we can obtain the optimal $\widehat{u}_{k}$ as
\begin{equation}\label{min:WMMSE:update u}
	\widehat{u}_{k}=\frac{\boldsymbol{f}_{k}^{\rm H}\boldsymbol{h}_{k}(\boldsymbol{r})}{\sum_{i=1}^{K}\left|\boldsymbol{h}_{k}^{\rm H}(\boldsymbol{r})\boldsymbol{f}_{i}\right|^{2}+\sigma_{n}^{2}}.
\end{equation}

\emph{3) Optimization of $\boldsymbol{w}$:} The subproblem to optimize $w_{k}$ can be expressed as
\begin{equation}
	\underset{w_{k}}{\min}~~w_{k}e_{k}-\mathrm{log}_{2}w_{k},
\end{equation}
which is convex in terms of $w_{k}$. By setting $\partial(w_{k}e_{k}-\mathrm{log}_{2}w_{k})/\partial w_{k}=0$, we can obtain the optimal $\widehat{w}_{k}$ as
\begin{equation}\label{min:WMMSE:update w}
	\widehat{w}_{k}=\frac{1}{e_{k}\mathrm{ln}2}.
\end{equation}

\emph{4) Optimization of $\boldsymbol{F}$:} The subproblem for optimizing $\boldsymbol{F}$ can be expressed as
\begin{subequations}\label{min:WMMSE:SubP F}
	\begin{align}
		\underset{\boldsymbol{F}}{\min}~~&\sum_{k=1}^{K}w_{k}e_{k}+\frac{1}{2\mu_{1}}\|\boldsymbol{F}-\boldsymbol{F}_{\rm RF}\boldsymbol{F}_{\rm BB}\|_{\rm F}^{2}\\
		\mathrm{s.t.}~~~&\eqref{AUX:PT constraint}.
	\end{align}
\end{subequations}
Introducing a dual variable $\nu_{1}\geq0$ associated with constraint~\eqref{AUX:PT constraint}, we define the Lagrangian function as
\begin{align}\label{min:WMMSE:SubP F:Lagrangian}
	L_{F,1}\left(\boldsymbol{F},\nu_{1}\right)\triangleq&\sum_{k=1}^{K}w_{k}e_{k}+\frac{1}{2\mu_{1}}\|\boldsymbol{F}-\boldsymbol{F}_{\rm RF}\boldsymbol{F}_{\rm BB}\|_{\rm F}^{2}\nonumber\\
	&+\nu_{1}\left(\|\boldsymbol{F}\|_{\rm F}^{2}-P_{\rm max}\right).
\end{align}
We define matrices $\boldsymbol{S}_{k}\in\mathbb{Z}^{N_{\rm T}\times N_{\rm T}K}$ and $\widetilde{\boldsymbol{H}}_{k}(\boldsymbol{r})\in\mathbb{C}^{N_{\rm T}K\times K}$, for $k=1,2,\cdots,K$, as
\begin{equation}\label{S_k:definition}
	\boldsymbol{S}_{k}\triangleq\big[\underbrace{\boldsymbol{0},\cdots,\boldsymbol{0}}_{k-1},\boldsymbol{I}_{N_{\rm T}},\underbrace{\boldsymbol{0},\cdots,\boldsymbol{0}}_{K-k}\big],
\end{equation}
\begin{equation}\label{widetilde H_k}
	\widetilde{\boldsymbol{H}}_{k}(\boldsymbol{r})\triangleq\left[\boldsymbol{S}_{1}^{\rm H}\boldsymbol{h}_{k}(\boldsymbol{r}),\boldsymbol{S}_{2}^{\rm H}\boldsymbol{h}_{k}(\boldsymbol{r}),\cdots,\boldsymbol{S}_{K}^{\rm H}\boldsymbol{h}_{k}(\boldsymbol{r})\right],
\end{equation}
respectively. Then, we can recast~\eqref{min:WMMSE:SubP F:Lagrangian} as~\eqref{min:WMMSE:SubP f:Lagrangian} shown at the top of this page, where we define $\boldsymbol{f}\triangleq\mathrm{vec}(\boldsymbol{F})$ and $\widetilde{\boldsymbol{f}}\triangleq\mathrm{vec}(\boldsymbol{F}_{\rm RF}\boldsymbol{F}_{\rm BB})$. Since~\eqref{min:WMMSE:SubP F} is a convex optimization problem and satisfies Slater's condition,
we can update $\widehat{\boldsymbol{f}}$ by~\eqref{min:WMMSE:update f} shown at the top of this page, where $\nu_{1}$ is obtained using the bisection method to satisfy the Karush-Kuhn-Tucker (KKT) conditions.

\setcounter{equation}{39}
\begin{figure*}
	\begin{align}\label{min:WMMSE:SubP f:Lagrangian}
		L_{f,1}\left(\boldsymbol{f},\nu_{1}\right)\triangleq\sum_{k=1}^{K}w_{k}\left(\left|u_{k}\right|^{2}\big(\|\widetilde{\boldsymbol{H}}_{k}^{\rm H}(\boldsymbol{r})\boldsymbol{f}\|_{2}^{2}+\sigma_{n}^{2}\big)+1-2\mathrm{Re}\{u_{k}\boldsymbol{h}_{k}^{\rm H}(\boldsymbol{r})\boldsymbol{S}_{k}\boldsymbol{f}\}\right)+\frac{1}{2\mu_{1}}\|\boldsymbol{f}-\widetilde{\boldsymbol{f}}\|_{2}^{2}+\nu_{1}\left(\|\boldsymbol{f}\|_{2}^{2}-P_{\rm max}\right).
	\end{align}
	\vspace*{-5pt}
\end{figure*}
\setcounter{equation}{41}

\setcounter{equation}{40}
\begin{figure*}[ht]
	\vspace*{-10pt}
	\begin{align}\label{min:WMMSE:update f}
		\widehat{\boldsymbol{f}}=\left(\sum_{k=1}^{K}w_{k}\left|u_{k}\right|^{2}\widetilde{\boldsymbol{H}}_{k}(\boldsymbol{r})\widetilde{\boldsymbol{H}}_{k}^{\rm H}(\boldsymbol{r})+\left(\frac{1}{2\mu_{1}}+\nu_{1}\right)\boldsymbol{I}_{N_{\rm T}K}\right)^{-1} \left(\sum_{k=1}^{K}w_{k}u_{k}^{\ast}\boldsymbol{S}_{k}^{\rm H}\boldsymbol{h}_{k}(\boldsymbol{r})+\frac{1}{2\mu_{1}}\widetilde{\boldsymbol{f}}\right).
	\end{align}
	\rule[0pt]{18.1cm}{0.05em}
	\vspace*{-15pt}
\end{figure*}
\setcounter{equation}{42}

\emph{5) Optimization of $\boldsymbol{F}_{\rm RF}$:} The subproblem to optimize $\boldsymbol{F}_{\rm RF}$ can be expressed as
\setcounter{equation}{41}
\begin{subequations}\label{min:WMMSE:SubP FRF}
	\begin{align}
		\underset{\boldsymbol{F}_{\rm RF}}{\min}~~&\|\boldsymbol{F}-\boldsymbol{F}_{\rm RF}\boldsymbol{F}_{\rm BB}\|_{\rm F}^{2}\\
		\mathrm{s.t.}~~~&\eqref{max:sum-rate:FRF constraint}.
	\end{align}
\end{subequations}
This problem is non-convex due to the constant-modulus constraint in~\eqref{max:sum-rate:FRF constraint}. Since the popular Riemannian manifold optimization method suffers from excessively high computational complexity, we respectively use two closed-form solutions for the FC-HBF and PC-HBF. Specifically, for the PC-HBF, the optimization of the diagonal blocks $\boldsymbol{f}_{1}^{\rm RF},\boldsymbol{f}_{2}^{\rm RF},\cdots,\boldsymbol{f}_{K}^{\rm RF}$ are mutually independent. Thus, we can obtain the optimal $\widehat{\boldsymbol{f}}_{k}^{\rm RF}$ by
\begin{equation}\label{min:WMMSE:update FRF PC}
	\widehat{\boldsymbol{f}}_{k}^{\rm RF}=\mathrm{exp}\left(j\angle\left(\left[\boldsymbol{F}\boldsymbol{F}_{\rm BB}^{\rm H}\right]_{(k-1)N_{\rm s}+1:kN_{\rm s},k}\right)\right).
\end{equation}
However, for the FC-HBF, no closed-form solution exists for the global optimum of $\boldsymbol{F}_{\rm RF}$. Fortunately, leveraging the near-orthogonal characteristic of the channels between the BS and each user, we obtain a suboptimal $\widehat{\boldsymbol{F}}_{\rm RF,FC}$ by~\cite{Yu2016Alternating}
\begin{equation}\label{min:WMMSE:update FRF FC}
	\widehat{\boldsymbol{F}}_{\rm RF,FC}=\mathrm{exp}\left(j\angle\left(\boldsymbol{F}\boldsymbol{F}_{\rm BB}^{\rm H}\right)\right).
\end{equation}

\emph{6) Optimization of $\boldsymbol{F}_{\rm BB}$:} The subproblem to optimize $\boldsymbol{F}_{\rm BB}$ is a least squares (LS) problem expressed as
\begin{equation}\label{min:WMMSE:SubP FBB}
	\underset{\boldsymbol{F}_{\rm BB}}{\min}~~\|\boldsymbol{F}-\boldsymbol{F}_{\rm RF}\boldsymbol{F}_{\rm BB}\|_{\rm F}^{2}.
\end{equation}
For the PC-HBF, we can readily use the LS solution to update $\boldsymbol{F}_{\rm BB}$ by
\begin{equation}\label{min:WMMSE:update FBB PC}
	\widehat{\boldsymbol{F}}_{\rm BB,PC}=\left(\boldsymbol{F}_{\rm RF}^{\rm H}\boldsymbol{F}_{\rm RF}\right)^{-1}\boldsymbol{F}_{\rm RF}^{\rm H}\boldsymbol{F}.
\end{equation}
Nevertheless, for the FC-HBF, we should ensure the orthogonality among the columns of $\boldsymbol{F}_{\rm BB}$ when updating $\boldsymbol{F}_{\rm RF}$ by~\eqref{min:WMMSE:update FRF FC}~\cite{Yu2016Alternating}. Thus, we obtain $\widehat{\boldsymbol{F}}_{\rm BB,FC}$ by
\begin{subequations}\label{min:WMMSE:update FBB FC}
	\begin{align}
		\widehat{\boldsymbol{F}}_{\rm BB,FC}=~&\boldsymbol{V}_{\rm BB}\boldsymbol{U}_{\rm BB}^{\rm H},\\
		\widehat{\boldsymbol{F}}_{\rm BB,FC}\gets~&\frac{\|\boldsymbol{F}\|_{\rm F}}{\|\boldsymbol{F}_{\rm RF}\widehat{\boldsymbol{F}}_{\rm BB,FC}\|_{\rm F}}\widehat{\boldsymbol{F}}_{\rm BB,FC},
	\end{align}
\end{subequations}
where $\boldsymbol{F}^{\rm H}\boldsymbol{F}_{\rm RF}\triangleq\boldsymbol{U}_{\rm BB}\boldsymbol{D}_{\rm BB}\boldsymbol{V}_{\rm BB}^{\rm H}$ is the singular value decomposition.

\emph{7) Update of $\mu_{1}$:} In the outer loop, we update $\mu_{1}$ by
\begin{equation}\label{min:WMMSE:update mu1}
	\mu_{1}\gets c_{1}\mu_{1}.
\end{equation}
Since $0<c_{1}<1$, the update of $\mu_{1}$ enhances the weight of the second term in~\eqref{L1:WMMSE+PDD}, thereby improving convergence performance.

Building upon the above descriptions, we summarize the complete procedure of the WMMSE-based PDD-HBF scheme in \textbf{Algorithm~\ref{Algorithm:PDD-HBF WMMSE}}.

\begin{algorithm}[!t]
	\caption{WMMSE-Based PDD-HBF Scheme}
	\label{Algorithm:PDD-HBF WMMSE}
	\begin{algorithmic}[1]
		\STATE \textbf{Input:} $N_{\rm T}$, $N_{\rm RF}$, $K$, $\boldsymbol{H}(\boldsymbol{r})$, $\sigma_{n}^{2}$, and $P_{\rm max}$.
		\STATE Initialize $\boldsymbol{F}$, $\boldsymbol{F}_{\rm RF}$, and $\boldsymbol{F}_{\rm BB}$ via~\eqref{min:WMMSE:initialize F}$\sim$\eqref{min:WMMSE:initialize FBB}.
		\REPEAT
		\REPEAT
		\STATE Update $\boldsymbol{u}$ via~\eqref{min:WMMSE:update u}.
		\STATE Update $\boldsymbol{w}$ via~\eqref{min:WMMSE:update w}.
		\STATE Update $\boldsymbol{F}$ via~\eqref{min:WMMSE:update f}, and obtain $\nu_{1}$ through the bisection method.
		\STATE Update $\boldsymbol{F}_{\rm RF}$ via~\eqref{min:WMMSE:update FRF PC} or~\eqref{min:WMMSE:update FRF FC}.
		\STATE Update $\boldsymbol{F}_{\rm BB}$ via~\eqref{min:WMMSE:update FBB PC} or~\eqref{min:WMMSE:update FBB FC}.
		\UNTIL{$R_{\rm SE}$ is converged.}
		\STATE Update $\mu_{1}$ via~\eqref{min:WMMSE:update mu1}.
		\UNTIL{$R_{\rm SE}$ is converged.}
		\STATE \textbf{Output:} $\boldsymbol{F}_{\rm RF}$, $\boldsymbol{F}_{\rm BB}$, and $R_{\rm SE}$.
	\end{algorithmic}
\end{algorithm}

\subsection{COD Method for RC Selection}\label{subsection:COD}
In this part, we use the COD method to optimize the RC selection. Specifically, we first initialize $\boldsymbol{r}$ according to the dimension-extended channel $\bar{\boldsymbol{H}}$ defined in~\eqref{hkbar:dimension-extended channel}. Then, we iteratively evaluate each selection by sequentially replacing each RC with another selectable RC candidate point, where the replacement will be made if the corresponding 
SE is improved. The detailed descriptions are presented as follows.

\emph{1) Initialization of $\boldsymbol{r}$:} According to the minimum index spacing $D_{\rm min}$, we define a candidate index set as
\begin{equation}
	\mathcal{C}_{\rm initial}\triangleq\left\lbrace c_{1},c_{2},\cdots,c_{N_{\rm C}}\right\rbrace,
\end{equation}
where we set $N_{\rm C}\triangleq\left\lceil N_{\rm EM}/D_{\rm min}\right\rceil\geq N_{\rm T}$ and $c_{n}=(n-1)D_{\rm min}+1$, for $n=1,2,\cdots,N_{\rm C}$. Then, we denote the candidate channel matrix by $\boldsymbol{H}_{\rm C}\in\mathbb{C}^{N_{\rm C}\times K}$, whose rows are selected from $\bar{\boldsymbol{H}}$ based on $\mathcal{C}_{\rm initial}$. To simply assess the channel conditions of the $N_{\rm C}$ candidate positions, we define the effective channel gain as
\begin{equation}
	g_{{\rm E},n}\triangleq\sum_{k=1}^{K}\left|\left[\boldsymbol{H}_{\rm C}\right]_{n,k}\right|,
\end{equation}
for $n=1,2,\cdots,N_{\rm C}$. Since a larger $g_{{\rm E},n}$ implies better channel condition, we initialize $\boldsymbol{r}$ as $\boldsymbol{r}^{(0)}$ according to the index set of $N_{\rm T}$ largest $g_{{\rm E},n}$, denoted by $\mathcal{X}^{(0)}$. As a result, $\boldsymbol{r}^{(0)}$ can be expressed as
\begin{align}\label{COD:initialize r}
	\big[\boldsymbol{r}^{(0)}\big]_{n}=\left\{\begin{aligned}
		1&,~~n\in\mathcal{X}^{(0)},\\
		0&,~~\mathrm{others}.
	\end{aligned}\right.
\end{align}

\emph{2) Iterative RC Optimization:} In the $p$th iteration, for $p\geq1$, we set a start point as $\boldsymbol{r}_{0}^{(p)}\gets\widehat{\boldsymbol{r}}^{(p-1)}$, where $\widehat{\boldsymbol{r}}^{(p-1)}$ denotes the optimal RC selection of the $(p-1)$th iteration. In particular, we set $\widehat{\boldsymbol{r}}^{(0)}\gets\boldsymbol{r}^{(0)}$. Additionally, we denote the index set of $\widehat{\boldsymbol{r}}^{(p)}$ by $\mathcal{X}^{(p)}$ based on \eqref{mapping:X to r}.

Then, for $m=\mathrm{mod}(p-1,N_{\rm T})+1$, we sequentially replace the $m$th element of $\mathcal{X}^{(p-1)}$ with $1,2,\cdots,N_{\rm EM}$ and generate a testing set for the $p$th iteration expressed as
\begin{equation}\label{Tp:testing set:generate}
	\mathcal{T}^{(p)}\triangleq\left\lbrace\boldsymbol{r}_{0}^{(p)}\right\rbrace\cup\left\lbrace\boldsymbol{r}_{1}^{(p)},\boldsymbol{r}_{2}^{(p)},\cdots,\boldsymbol{r}_{N_{\rm EM}}^{(p)}\right\rbrace,
\end{equation}
where $\boldsymbol{r}_{n}^{(p)}$ is obtained according to the mapping relationship defined in~\eqref{mapping:X to r}, for $n=1,2,\cdots,N_{\rm EM}$. To satisfy the RC selection constraint~\eqref{max:sum-rate:RC constraint}, we have
\begin{equation}\label{Tp:testing set:pruning}
	\mathcal{T}^{(p)}\gets\mathcal{T}^{(p)}\setminus\boldsymbol{r}_{n}^{(p)},~~\mathrm{if}~\boldsymbol{r}_{n}^{(p)}\notin\mathcal{A}_{r}.
\end{equation}

Next, we substitute each $\boldsymbol{r}_{n}^{(p)}\in\mathcal{T}^{(p)}$ into~\eqref{max:sum-rate AUX} and solve it using the WMMSE-based PDD-HBF scheme described in \textbf{Algorithm~\ref{Algorithm:PDD-HBF WMMSE}}. The resulting SE, denoted by $R_{\rm SE}(\boldsymbol{r}_{n}^{(p)})$, is calculated based on~\eqref{Rsum:sum-rate}. As a result, the optimal RCs of the $p$th iteration can be obtained as
\begin{equation}\label{COD:obtain r}
	\widehat{\boldsymbol{r}}^{(p)}=\arg\max_{\boldsymbol{r}^{(p)}\in\mathcal{T}^{(p)}}R_{\rm SE}\big(\boldsymbol{r}^{(p)}\big).
\end{equation}

\emph{3) Stopping Condition:} The stopping condition of the COD method can be expressed as
\begin{equation}\label{COD:stop condition}
	\widehat{\boldsymbol{r}}^{(p)}=\widehat{\boldsymbol{r}}^{(p-N_{\rm T})},~~p\geq N_{\rm T}+1,
\end{equation}
which indicates that any replacement of RCs cannot further improve the SE. Note that each iteration of the COD method ensures the SE is monotonically non-decreasing, thereby guaranteeing the convergence.

Based on the WMMSE-based PDD-HBF scheme and the COD method, the complete TLAO scheme to solve~\eqref{max:sum-rate} for SE maximization (TLAO-SE) is summarized in \textbf{Algorithm~\ref{Algorithm:TLAO THB SR}}.

\begin{algorithm}[!t]
	\caption{TLAO-SE Scheme}
	\label{Algorithm:TLAO THB SR}
	\begin{algorithmic}[1]
		\STATE \textbf{Input:} $N_{\rm EM}$, $N_{\rm T}$, $N_{\rm RF}$, $K$, $\bar{\boldsymbol{H}}$, $D_{\rm min}$, $\sigma_{n}^{2}$, and $P_{\rm max}$.
		\STATE Initialize $\boldsymbol{r}^{(0)}$ via~\eqref{COD:initialize r}. Set $\widehat{\boldsymbol{r}}^{(0)}\gets\boldsymbol{r}^{(0)}$ and $p\gets0$.
		\REPEAT
		\STATE Set $p\gets p+1$.
		\STATE Generate $\mathcal{T}^{(p)}$ via~\eqref{Tp:testing set:generate} and~\eqref{Tp:testing set:pruning}.
		\STATE Evaluate each element in $\mathcal{T}^{(p)}$ by solving~\eqref{max:sum-rate AUX} using \textbf{Algorithm~\ref{Algorithm:PDD-HBF WMMSE}}.
		\STATE Obtain $\widehat{\boldsymbol{r}}^{(p)}$ via~\eqref{COD:obtain r}.
		\UNTIL{Conditions in~\eqref{COD:stop condition} is satisfied.}
		\STATE \textbf{Output:} $\boldsymbol{r}$, $\boldsymbol{F}_{\rm RF}$, $\boldsymbol{F}_{\rm BB}$, and $R_{\rm SE}$.
	\end{algorithmic}
\end{algorithm}

\section{THBF Design for EE Maximization}\label{section:THB EE}
\subsection{Problem Formulation}
Based on the definition of $\eta_{\rm EE}$ in~\eqref{etaEE:definition}, the EE maximization problem for the THBF design can be formulated as
\begin{subequations}\label{max:EE}
	\begin{align}
		\underset{\boldsymbol{r},\boldsymbol{F}_{\rm RF},\boldsymbol{F}_{\rm BB}}{\max}~~&\frac{\sum_{k=1}^{K}\mathrm{log}_{2}\left(1+\gamma_{k}\right)}{\|\boldsymbol{F}_{\rm RF}\boldsymbol{F}_{\rm BB}\|_{\rm F}^{2}/\eta_{\rm PA}+P_{\rm C}}\label{max:EE:objective}\\
		\mathrm{s.t.}~~~~~~&\boldsymbol{r}\in\mathcal{A}_{r},\\
		&\boldsymbol{F}_{\rm RF}\in\mathcal{F}_{\rm FC}~\mathrm{or}~\mathcal{F}_{\rm PC},\label{max:EE:FRF constraint}\\
		&\|\boldsymbol{F}_{\rm RF}\boldsymbol{F}_{\rm BB}\|_{\rm F}^{2}\leq P_{\rm max},\label{max:EE:PT constraint}
	\end{align}
\end{subequations}
where $P_{\rm C}$ can be expressed as
\begin{equation}
	P_{\rm C}\triangleq P_{\rm LO}+N_{\rm RF}\left(P_{\rm RF}+2P_{\rm DAC}\right)+N_{\rm PS}P_{\rm PS}.
\end{equation}

Since the TLAO framework proposed in Section~\ref{section:THB SR} can be readily used for solving~\eqref{max:EE}, we focus on the HBF design with fixed RCs in this section, i.e.,
\begin{subequations}\label{max:EE:fixed RCs}
	\begin{align}
		\underset{\boldsymbol{F}_{\rm RF},\boldsymbol{F}_{\rm BB}}{\max}~~&\frac{\sum_{k=1}^{K}\mathrm{log}_{2}\left(1+\gamma_{k}\right)}{\|\boldsymbol{F}_{\rm RF}\boldsymbol{F}_{\rm BB}\|_{\rm F}^{2}/\eta_{\rm PA}+P_{\rm C}}\label{max:EE:fixed RCs:objective}\\
		\mathrm{s.t.}~~~~~&~\eqref{max:EE:FRF constraint}~\mathrm{and}~\eqref{max:EE:PT constraint}.
	\end{align}
\end{subequations}
Similar to~\eqref{max:sum-rate AUX} and~\eqref{min:WMMSE PDD}, we handle the coupled variables $\boldsymbol{F}_{\rm RF}$ and $\boldsymbol{F}_{\rm BB}$ by introducing the auxiliary variable $\boldsymbol{F}\triangleq\boldsymbol{F}_{\rm RF}\boldsymbol{F}_{\rm BB}$ and employing the PDD method. Then,we can transform~\eqref{max:EE:fixed RCs} into
\begin{subequations}\label{max:EE:PDD}
	\begin{align}
		\underset{\boldsymbol{F},\boldsymbol{F}_{\rm RF},\boldsymbol{F}_{\rm BB}}{\max}~~&L_{2}\left(\boldsymbol{F},\boldsymbol{F}_{\rm RF},\boldsymbol{F}_{\rm BB}\right)\label{max:EE:PDD:objective}\\
		\mathrm{s.t.}~~~~~~&\|\boldsymbol{F}\|_{\rm F}^{2}\leq P_{\rm max}~\mathrm{and}~\eqref{max:EE:FRF constraint}.\label{max:EE:PDD:constraints}
	\end{align}
\end{subequations}
We can express $L_{2}\left(\boldsymbol{F},\boldsymbol{F}_{\rm RF},\boldsymbol{F}_{\rm BB}\right)$ as
\begin{align}
	&L_{2}\left(\boldsymbol{F},\boldsymbol{F}_{\rm RF},\boldsymbol{F}_{\rm BB}\right)\nonumber\\
	\triangleq~&\frac{\sum_{k=1}^{K}\mathrm{log}_{2}\left(1+\widetilde{\gamma}_{k}\right)}{\|\boldsymbol{F}\|_{\rm F}^{2}/\eta_{\rm PA}+P_{\rm C}}-\frac{1}{2\mu_{2}}\|\boldsymbol{F}-\boldsymbol{F}_{\rm RF}\boldsymbol{F}_{\rm BB}\|_{\rm F}^{2},
\end{align}
where $\mu_{2}>0$ is a penalty coefficient, and $\widetilde{\gamma}_{k}$ is defined in~\eqref{gamma_k widetilde}, for $k=1,2,\cdots,K$. In the following, we will present the DQTFP and LDTFP schemes for solving~\eqref{max:EE:PDD}.

\subsection{DQTFP Scheme for HBF Design}
The primary challenge in solving~\eqref{max:EE:PDD} is the inherent nested fractional form of the first term in~\eqref{max:EE:PDD:objective}, including the outer EE ratio and the inner SINR ratio. We use the quadratic transform (QT) to decouple the outer EE ratio and define $G_{q}\left(\boldsymbol{F},\boldsymbol{F}_{\rm RF},\boldsymbol{F}_{\rm BB},\rho\right)$ as~\cite{Shen2018Fractional}
\begin{align}\label{Gq:QT}
	&G_{q}\left(\boldsymbol{F},\boldsymbol{F}_{\rm RF},\boldsymbol{F}_{\rm BB},\rho\right)\nonumber\\
	\triangleq~&2\rho\sqrt{{\textstyle\sum_{k=1}^{K}}\mathrm{log}_{2}\left(1+\widetilde{\gamma}_{k}\right)}-\rho^{2}\left(\|\boldsymbol{F}\|_{\rm F}^{2}/\eta_{\rm PA}+P_{\rm C}\right)\nonumber\\
	&-\frac{1}{2\mu_{2}}\|\boldsymbol{F}-\boldsymbol{F}_{\rm RF}\boldsymbol{F}_{\rm BB}\|_{\rm F}^{2},
\end{align}
where $\rho$ is an auxiliary variable. 

Then, problem~\eqref{max:EE:PDD} can be equivalently converted to
\begin{subequations}\label{max:EE:QT}
	\begin{align}
		\underset{\boldsymbol{F},\boldsymbol{F}_{\rm RF},\boldsymbol{F}_{\rm BB},\rho}{\max}~~&G_{q}\left(\boldsymbol{F},\boldsymbol{F}_{\rm RF},\boldsymbol{F}_{\rm BB},\rho\right)\\
		\mathrm{s.t.}~~~~~~~&\|\boldsymbol{F}\|_{\rm F}^{2}\leq P_{\rm max},~ \rho\in\mathbb{R},~\mathrm{and}~\eqref{max:EE:FRF constraint},
	\end{align}
\end{subequations}
which can be decomposed into multiple subproblems and solved iteratively. Specifically, since $G_{q}$ is concave in terms of $\rho$, we can update $\rho$ by setting $\partial G_{q}/\partial\rho=0$, i.e.,
\begin{equation}\label{max:EE:DQTFP:update rho}
	\widehat{\rho}\gets\frac{\sqrt{{\textstyle\sum_{k=1}^{K}}\mathrm{log}_{2}\left(1+\widetilde{\gamma}_{k}\right)}}{\|\boldsymbol{F}\|_{\rm F}^{2}/\eta_{\rm PA}+P_{\rm C}}.
\end{equation}
However, the subproblem for optimizing $\boldsymbol{F}$ is still difficult to solve due to the inner fractional form of $\widetilde{\gamma}_{k}$. To mitigate this challenge, we further employ QT for each $\widetilde{\gamma}_{k}$ and introduce an auxiliary vector $\boldsymbol{\tau}\triangleq[\tau_{1},\tau_{2},\cdots,\tau_{K}]^{\rm T}$. Based on $G_{q}$, we define $G_{qq}\left(\boldsymbol{f},\boldsymbol{F}_{\rm RF},\boldsymbol{F}_{\rm BB},\rho,\boldsymbol{\tau}\right)$ as \eqref{Gqq:DQT} shown at the top of this page, where we replace $\boldsymbol{F}$ with $\boldsymbol{f}\triangleq\mathrm{vec}(\boldsymbol{F})$, and the matrices $\boldsymbol{S}_{k}$ and $\widetilde{\boldsymbol{H}}_{k}$ are defined in \eqref{S_k:definition} and \eqref{widetilde H_k}, respectively, for $k=1,2,\cdots,K$. As a result, we can equivalently transform \eqref{max:EE:QT} into
\setcounter{equation}{64}
\begin{subequations}\label{max:EE:DQTFP}
	\begin{align}
		\underset{\boldsymbol{f},\boldsymbol{F}_{\rm RF},\boldsymbol{F}_{\rm BB},\rho,\boldsymbol{\tau}}{\max}~~&G_{qq}\left(\boldsymbol{f},\boldsymbol{F}_{\rm RF},\boldsymbol{F}_{\rm BB},\rho,\boldsymbol{\tau}\right)\label{max:EE:DQTFP:objective}\\
		\mathrm{s.t.}~~~~~~~~&\|\boldsymbol{f}\|_{2}^{2}\leq P_{\rm max},~\rho\in\mathbb{R},~\boldsymbol{\tau}\in\mathbb{R}^{K},\nonumber\\
		&\mathrm{and}~\eqref{max:EE:FRF constraint}.
	\end{align}
\end{subequations}
Since the original problem \eqref{max:EE:PDD} is converted to \eqref{max:EE:DQTFP} based on two sequential QT operations, we refer to this HBF design as the DQTFP scheme. In the following, we will describe the detailed procedures to solve~\eqref{max:EE:DQTFP} by alternately optimizing $\boldsymbol{\tau}$, $\rho$, $\boldsymbol{f}$, $\boldsymbol{F}_{\rm RF}$, and $\boldsymbol{F}_{\rm BB}$ based on the DQTFP scheme.

\setcounter{equation}{63}
\begin{figure*}
	\begin{align}\label{Gqq:DQT}
		G_{qq}\left(\boldsymbol{f},\boldsymbol{F}_{\rm RF},\boldsymbol{F}_{\rm BB},\rho,\boldsymbol{\tau}\right)\triangleq~& 2\rho\sqrt{{\textstyle\sum_{k=1}^{K}}\mathrm{log}_{2}\left(1+2\mathrm{Re}\left\lbrace\tau_{k}^{\ast}\boldsymbol{h}_{k}^{\rm H}(\boldsymbol{r})\boldsymbol{S}_{k}\boldsymbol{f}\right\rbrace-\left|\tau_{k}\right|^{2}\left(\|\widetilde{\boldsymbol{H}}_{k}^{\rm H}(\boldsymbol{r})\boldsymbol{f}\|_{2}^{2}-\left|\boldsymbol{h}_{k}^{\rm H}(\boldsymbol{r})\boldsymbol{S}_{k}\boldsymbol{f}\right|^{2}+\sigma_{n}^{2}\right)\right)}\nonumber\\
		&-\rho^{2}\left(\|\boldsymbol{f}\|_{2}^{2}/\eta_{\rm PA}+P_{\rm C}\right)-\frac{1}{2\mu_{2}}\|\boldsymbol{f}-\mathrm{vec}\left(\boldsymbol{F}_{\rm RF}\boldsymbol{F}_{\rm BB}\right)\|_{2}^{2}.
	\end{align}
	\rule[0pt]{18.1cm}{0.05em}
	\vspace*{-15pt}
\end{figure*}
\setcounter{equation}{65}

\emph{1) Optimization of $\boldsymbol{\tau}$:} Due to the independence of $\tau_{1}$, $\tau_{2}$, $\cdots$, $\tau_{K}$ in~\eqref{max:EE:DQTFP}, the subproblem for optimizing $\tau_{k}$ can be expressed as
\begin{align}
	\underset{\tau_{k}}{\max}~~&2\mathrm{Re}\left\lbrace\tau_{k}^{\ast}\boldsymbol{h}_{k}^{\rm H}(\boldsymbol{r})\boldsymbol{S}_{k}\boldsymbol{f}\right\rbrace\nonumber\\
	-&\left|\tau_{k}\right|^{2}\left(\|\widetilde{\boldsymbol{H}}_{k}^{\rm H}(\boldsymbol{r})\boldsymbol{f}\|_{2}^{2}-\left|\boldsymbol{h}_{k}^{\rm H}(\boldsymbol{r})\boldsymbol{S}_{k}\boldsymbol{f}\right|^{2}+\sigma_{n}^{2}\right),
\end{align}
which is concave in terms of $\tau_{k}$. Then, the optimal $\widehat{\tau}_{k}$ can be obtained by
\begin{equation}\label{max:EE:DQTFP:update tau}
	\widehat{\tau}_{k}=\frac{\boldsymbol{h}_{k}^{\rm H}(\boldsymbol{r})\boldsymbol{S}_{k}\boldsymbol{f}}{\|\widetilde{\boldsymbol{H}}_{k}^{\rm H}(\boldsymbol{r})\boldsymbol{f}\|_{2}^{2}-\left|\boldsymbol{h}_{k}^{\rm H}(\boldsymbol{r})\boldsymbol{S}_{k}\boldsymbol{f}\right|^{2}+\sigma_{n}^{2}}.
\end{equation}

\emph{2) Optimization of $\rho$:} 
Since the multi-ratio QT operation keeps the values of $G_{q}$ and $G_{qq}$ unchanged when all the variables are jointly optimized~\cite{Shen2018Fractional}, we employ~\eqref{max:EE:DQTFP:update rho} to update $\rho$ by setting $\partial G_{q}/\partial\rho=0$.

\emph{3) Optimization of $\boldsymbol{f}$:} The subproblem for optimizing $\boldsymbol{f}$ can be expressed as
\begin{subequations}\label{max:EE:SubP f}
	\begin{align}
		\underset{\boldsymbol{f}}{\max}~~&2\rho\sqrt{{\textstyle\sum_{k=1}^{K}}\mathrm{log}_{2}\left(1+g_{f,k}(\boldsymbol{f})\right)}\nonumber\\
		-&\rho^{2}\left(\|\boldsymbol{f}\|_{2}^{2}/\eta_{\rm PA}+P_{\rm C}\right)-\frac{1}{2\mu_{2}}\|\boldsymbol{f}-\widetilde{\boldsymbol{f}}\|_{2}^{2}\label{max:EE:SubP f:objective}\\
		\mathrm{s.t.}~~~&\|\boldsymbol{f}\|_{2}^{2}\leq P_{\rm max},\label{max:EE:SubP f:PT constraint}
	\end{align}
\end{subequations}
where we define $\widetilde{\boldsymbol{f}}\triangleq\mathrm{vec}(\boldsymbol{F}_{\rm RF}\boldsymbol{F}_{\rm BB})$ and express $g_{f,k}(\boldsymbol{f})$ as
\begin{align}
	g_{f,k}(\boldsymbol{f})&=2\mathrm{Re}\left\lbrace\tau_{k}^{\ast}\boldsymbol{h}_{k}^{\rm H}(\boldsymbol{r})\boldsymbol{S}_{k}\boldsymbol{f}\right\rbrace\nonumber\\
	&-\left|\tau_{k}\right|^{2}\left({\textstyle\sum_{i=1,i\neq k}^{K}}\left|\boldsymbol{h}_{i}^{\rm H}(\boldsymbol{r})\boldsymbol{S}_{i}\boldsymbol{f}\right|^{2}+\sigma_{n}^{2}\right).
\end{align}
Since the Hessian matrix of $g_{f,k}(\boldsymbol{f})$ is calculated as
\begin{equation}
	\nabla^{2}g_{f,k}(\boldsymbol{f})=-\left|\tau_{k}\right|^{2}\sum_{i=1,i\neq k}^{K}\boldsymbol{S}_{i}^{\rm H}\boldsymbol{h}_{i}(\boldsymbol{r})\boldsymbol{h}_{k}^{\rm H}(\boldsymbol{r})\boldsymbol{S}_{i}\preceq\boldsymbol{0},
\end{equation}
it can be seen that $g_{f,k}(\boldsymbol{f})$ is concave in terms of $\boldsymbol{f}$. Besides, since the functions of $\sqrt{x}$ and $\mathrm{log}_{2}x$ are both concave and non-decreasing, then the first term of~\eqref{max:EE:SubP f:objective} is concave in terms of $\boldsymbol{f}$. 
As a result, \eqref{max:EE:SubP f} is a convex problem and can be solved using the CVX toolbox.

\emph{4) Optimization of $\boldsymbol{F}_{\rm RF}$ and $\boldsymbol{F}_{\rm BB}$:} The subproblems to optimize $\boldsymbol{F}_{\rm RF}$ and $\boldsymbol{F}_{\rm BB}$ are the same as~\eqref{min:WMMSE:SubP FRF} and~\eqref{min:WMMSE:SubP FBB}, respectively. Therefore, we can readily employ~\eqref{min:WMMSE:update FRF PC} or~\eqref{min:WMMSE:update FRF FC} to update $\boldsymbol{F}_{\rm RF}$, and employ~\eqref{min:WMMSE:update FBB PC} or~\eqref{min:WMMSE:update FBB FC} to update $\boldsymbol{F}_{\rm BB}$.

Based on the aforementioned procedures, we summarize the DQTFP scheme for the HBF design with fixed RCs in \textbf{Algorithm~\ref{Algorithm:DQTFP}}.

\begin{algorithm}[!t]
	\caption{DQTFP Scheme}
	\label{Algorithm:DQTFP}
	\begin{algorithmic}[1]
		\STATE \textbf{Input:} $N_{\rm T}$, $N_{\rm RF}$, $K$, $\sigma_{n}^{2}$, $P_{\rm max}$, $P_{\rm C}$, and $\eta_{\rm PA}$.
		\STATE Initialize $\boldsymbol{F}$, $\boldsymbol{F}_{\rm RF}$, and $\boldsymbol{F}_{\rm BB}$ via~\eqref{min:WMMSE:initialize F}$\sim$\eqref{min:WMMSE:initialize FBB}.
		\REPEAT
		\REPEAT
		\STATE Update $\boldsymbol{\tau}$ via~\eqref{max:EE:DQTFP:update tau}.
		\STATE Update $\rho$ via~\eqref{max:EE:DQTFP:update rho}.
		\STATE Update $\boldsymbol{F}$ by solving~\eqref{max:EE:SubP f}.\label{Algorithm:DQTFP:update f}
		\STATE Update $\boldsymbol{F}_{\rm RF}$ via~\eqref{min:WMMSE:update FRF PC} or~\eqref{min:WMMSE:update FRF FC}.
		\STATE Update $\boldsymbol{F}_{\rm BB}$ via~\eqref{min:WMMSE:update FBB PC} or~\eqref{min:WMMSE:update FBB FC}.
		\UNTIL{$\eta_{\rm EE}$ is converged.}
		\STATE Update $\mu_{2}$ as $\mu_{2}\gets\mu_{2}c_{2}$, where we set $0<c_{2}<1$.
		\UNTIL{$\eta_{\rm EE}$ is converged.}
		\STATE \textbf{Output:} $\boldsymbol{F}_{\rm RF}$, $\boldsymbol{F}_{\rm BB}$, and $\eta_{\rm EE}$.
	\end{algorithmic}
\end{algorithm}

\subsection{LDTFP Scheme for HBF Design}
The dominant computational complexity of \textbf{Algorithm~\ref{Algorithm:DQTFP}} stems from step~\ref{Algorithm:DQTFP:update f}, where we have to use the CVX toolbox to solve \eqref{max:EE:SubP f}. Consequently, the computational complexity is high when iteratively optimizing the HBF design and RC selection under the TLAO framework. In this subsection, we develop a low-complexity algorithm to design the analog and digital beamformers with fixed RCs.

We first review the nested fractional form in~\eqref{max:EE:PDD:objective}. Note that the outer EE expression of the first term exhibits a single-ratio fractional form, where both Dinkelbach's transform (DBT) and QT can be used to decouple it. Since the DBT can avoid the existence of $\sqrt{x}$ function and simplify the optimization, we employ the DBT to transform \eqref{max:EE:PDD:objective} into
\begin{subequations}\label{max:EE:DBT}
	\begin{align}
		\underset{\boldsymbol{F},\boldsymbol{F}_{\rm RF},\boldsymbol{F}_{\rm BB}}{\max}~~&\sum_{k=1}^{K}\mathrm{log}_{2}\left(1+\widetilde{\gamma}_{k}\right)-\omega\left(\|\boldsymbol{F}\|_{\rm F}^{2}/\eta_{\rm PA}+P_{\rm C}\right)\nonumber\\
		&-\frac{1}{2\mu_{2}}\|\boldsymbol{F}-\boldsymbol{F}_{\rm RF}\boldsymbol{F}_{\rm BB}\|_{\rm F}^{2},\\
		\mathrm{s.t.}~~~~~~&\|\boldsymbol{F}\|_{\rm F}^{2}\leq P_{\rm max}~\mathrm{and}~\eqref{max:EE:FRF constraint},
	\end{align}
\end{subequations}
where $\omega$ is an auxiliary variable and can be iteratively updated by~\cite{Shen2018Fractional}
\begin{equation}\label{LDTFP:update omega}
	\widehat{\omega}=\frac{{\textstyle\sum_{k=1}^{K}\left(1+\widetilde{\gamma}_{k}\right)}}{\|\boldsymbol{F}\|_{\rm F}^{2}/\eta_{\rm PA}+P_{\rm C}}.
\end{equation}

Then, we formulate the epigraph form of problem~\eqref{max:EE:DBT} as
\begin{subequations}\label{max:EE:DBT+epi}
	\begin{align}
		\underset{\boldsymbol{F},\boldsymbol{F}_{\rm RF},\boldsymbol{F}_{\rm BB},\boldsymbol{t}}{\max}~~&\sum_{k=1}^{K}\mathrm{log}_{2}\left(1+t_{k}\right)-\omega\left(\|\boldsymbol{F}\|_{\rm F}^{2}/\eta_{\rm PA}+P_{\rm C}\right)\label{max:EE:DBT+epi:objective}\nonumber\\
		&-\frac{1}{2\mu_{2}}\|\boldsymbol{F}-\boldsymbol{F}_{\rm RF}\boldsymbol{F}_{\rm BB}\|_{\rm F}^{2}\\
		\mathrm{s.t.}~~~~~~~&t_{k}\leq\widetilde{\gamma}_{k} ,\label{max:EE:DBT+epi:epi constraint}\\
		&\|\boldsymbol{F}\|_{\rm F}^{2}\leq P_{\rm max}~\mathrm{and}~\eqref{max:EE:FRF constraint},
	\end{align}
\end{subequations}
where $\boldsymbol{t}\triangleq[t_{1},t_{2},\cdots,t_{K}]^{\rm T}\in\mathbb{R}^{K}$ is an auxiliary vector. We can alternately optimize $\boldsymbol{t}$ and $\{\boldsymbol{F},\boldsymbol{F}_{\rm RF},\boldsymbol{F}_{\rm BB}\}$ to solve this problem. 

We consider the subproblem for optimizing $\boldsymbol{t}$, i.e.,
\begin{subequations}\label{max:EE:DBT+epi:SubP t}
	\begin{align}
		\underset{\boldsymbol{t}}{\max}~~&\sum_{k=1}^{K}\mathrm{log}_{2}\left(1+t_{k}\right)\\
		\mathrm{s.t.}~~~&t_{k}\leq\widetilde{\gamma}_{k},
	\end{align}
\end{subequations}
where the optimal $\widehat{t}_{k}$ is
\begin{equation}\label{update t:strong duality}
	\widehat{t}_{k}=\widetilde{\gamma}_{k},
\end{equation}
since $\mathrm{log}_{2}(x)$ is non-decreasing. Then, the Lagrangian function of~\eqref{max:EE:DBT+epi:SubP t}, denoted as $L_{t}(\boldsymbol{t},\boldsymbol{\lambda})$, can be expressed as
\begin{equation}\label{L_t:Lagrangian}
	L_{t}(\boldsymbol{t},\boldsymbol{\lambda})=\sum_{k=1}^{K}\mathrm{log}_{2}\left(1+t_{k}\right)-\sum_{k=1}^{K}\lambda_{k}(t_{k}-\widetilde{\gamma}_{k}),
\end{equation}
where $\boldsymbol{\lambda}\triangleq[\lambda_{1},\lambda_{2},\cdots,\lambda_{K}]^{\rm T}\succeq\boldsymbol{0}$ is the dual vector. Since problem~\eqref{max:EE:DBT+epi:SubP t} is convex and satisfies strong duality, the optimization of the original problem is equivalent to that of the dual problem expressed as
\begin{equation}
	\underset{\boldsymbol{\lambda}\succeq\boldsymbol{0}}{\min}~\underset{\boldsymbol{t}}{\sup}~~L_{t}\left(\boldsymbol{t},\boldsymbol{\lambda}\right).
\end{equation}
According to the KKT conditions, the optimal $(\widehat{\boldsymbol{t}},\widehat{\boldsymbol{\lambda}})$ should satisfy $\nabla_{\boldsymbol{t}} L_{t}(\widehat{\boldsymbol{t}},\widehat{\boldsymbol{\lambda}})=\boldsymbol{0}$. Then, we have
\begin{equation}\label{widetilde lambda_k:KKT}
	\widehat{\lambda}_{k}=\frac{1}{\left(1+\widehat{t}_{k}\right)\mathrm{ln}2}=\frac{1}{\left(1+\widetilde{\gamma}_{k}\right)\mathrm{ln}2}.
\end{equation}
Substituting~\eqref{widetilde lambda_k:KKT} into~\eqref{L_t:Lagrangian}, we have
\begin{align}\label{L_t:t lambdaHat}
	&L_{t}(\boldsymbol{t},\widehat{\boldsymbol{\lambda}})=\sum_{k=1}^{K}\mathrm{log}_{2}\left(1+t_{k}\right)-\sum_{k=1}^{K}\frac{1}{\left(1+\widetilde{\gamma}_{k}\right)\mathrm{ln}2}\left(t_{k}-\widetilde{\gamma}_{k}\right)\nonumber\\
	&=\sum_{k=1}^{K}\left(\mathrm{log}_{2}\left(1+t_{k}\right)-\frac{t_{k}}{\mathrm{ln}2}+\frac{\left(t_{k}+1\right)\widetilde{\gamma}_{k}}{\left(1+\widetilde{\gamma}_{k}\right)\mathrm{ln}2}\right).
\end{align}
Then, we define $G_{l}$ as~\eqref{Gl} shown at the top of next page by substituting \eqref{L_t:t lambdaHat} and $\boldsymbol{f}\triangleq\mathrm{vec}(\boldsymbol{F})$ into \eqref{max:EE:DBT+epi:objective}. As a result, based on the Lagrange dual transform, we can convert~\eqref{max:EE:DBT+epi:objective} into
\setcounter{equation}{80}
\begin{equation}\label{max:EE:LDT:objective}
	\underset{\boldsymbol{f},\boldsymbol{F}_{\rm RF},\boldsymbol{F}_{\rm BB},\boldsymbol{t}}{\max}~~G_{l}\left(\boldsymbol{f},\boldsymbol{F}_{\rm RF},\boldsymbol{F}_{\rm BB},\boldsymbol{t}\right).
\end{equation}

\setcounter{equation}{79}
\begin{figure*}
	\begin{align}\label{Gl}
		G_{l}\left(\boldsymbol{f},\boldsymbol{F}_{\rm RF},\boldsymbol{F}_{\rm BB},\boldsymbol{t}\right)\triangleq&\sum_{k=1}^{K}\left(\mathrm{log}_{2}\left(1+t_{k}\right)-\frac{t_{k}}{\mathrm{ln}2}+\frac{\left(t_{k}+1\right)\left|\boldsymbol{h}_{k}^{\rm H}(\boldsymbol{r})\boldsymbol{S}_{k}\boldsymbol{f}\right|^{2}}{\mathrm{ln}2\left(\|\widetilde{\boldsymbol{H}}_{k}^{\rm H}(\boldsymbol{r})\boldsymbol{f}\|_{2}^{2}+\sigma_{n}^{2}\right)}\right)-\omega\left(\|\boldsymbol{f}\|_{2}^{2}/\eta_{\rm PA}+P_{\rm C}\right)\nonumber\\
		&-\frac{1}{2\mu_{2}}\|\boldsymbol{f}-\mathrm{vec}(\boldsymbol{F}_{\rm RF}\boldsymbol{F}_{\rm BB})\|_{2}^{2}.
	\end{align}
	\rule[0pt]{18.1cm}{0.05em}
\end{figure*}
\setcounter{equation}{81}

To decouple the multi-ratio fractional form, we employ the QT operation and define $G_{lq}$ as~\eqref{Glq} shown at the top of next page. As a result, the EE maximization problem with fixed RCs is finally transformed into
\setcounter{equation}{82}
\begin{subequations}\label{max:EE:LDT+QT}
	\begin{align}
		\underset{\boldsymbol{f},\boldsymbol{F}_{\rm RF},\boldsymbol{F}_{\rm BB},\boldsymbol{t},\boldsymbol{z}}{\max}~~&G_{lq}\left(\boldsymbol{f},\boldsymbol{F}_{\rm RF},\boldsymbol{F}_{\rm BB},\boldsymbol{t},\boldsymbol{z}\right)\\
		\mathrm{s.t.}~~~~~~~~&\|\boldsymbol{f}\|_{2}^{2}\leq P_{\rm max}~\mathrm{and}~\eqref{max:EE:FRF constraint}.
	\end{align}
\end{subequations}

\setcounter{equation}{81}
\begin{figure*}
	\vspace*{-5pt}
	\begin{align}\label{Glq}
		G_{lq}\left(\boldsymbol{f},\boldsymbol{F}_{\rm RF},\boldsymbol{F}_{\rm BB},\boldsymbol{t},\boldsymbol{z}\right)\triangleq&\sum_{k=1}^{K}\left(\mathrm{log}_{2}\left(1+t_{k}\right)-\frac{t_{k}}{\mathrm{ln}2}+2\sqrt{t_{k}+1}\mathrm{Re}\left\lbrace z_{k}^{\ast}\boldsymbol{h}_{k}^{\rm H}(\boldsymbol{r})\boldsymbol{S}_{k}\boldsymbol{f}\right\rbrace-\left|z_{k}\right|^{2}\mathrm{ln}2\left(\|\widetilde{\boldsymbol{H}}_{k}^{\rm H}(\boldsymbol{r})\boldsymbol{f}\|_{2}^{2}+\sigma_{n}^{2}\right)\right)\nonumber\\
		&-\omega\left(\|\boldsymbol{f}\|_{2}^{2}/\eta_{\rm PA}+P_{\rm C}\right)-\frac{1}{2\mu_{2}}\|\boldsymbol{f}-\mathrm{vec}(\boldsymbol{F}_{\rm RF}\boldsymbol{F}_{\rm BB})\|_{2}^{2}.
	\end{align}
	\rule[0pt]{18.1cm}{0.05em}
	\vspace*{-15pt}
\end{figure*}
\setcounter{equation}{83}

In the following, we will provide the details to solve~\eqref{max:EE:LDT+QT} based on the LDTFP scheme.

\emph{1) Optimization of $\boldsymbol{t}$:} 
Since it is challenging to obtain a closed-form solution by setting $\partial G_{lq}/\partial t_{k}=0$ with initialized~$\boldsymbol{z}$, we equivalently update $t_{k}$ by setting $\partial G_{l}/\partial t_{k}=0$ based on the unchanged value and first-order derivation of the QT operation. Then, the optimal $\widehat{t}_{k}$ can be calculated as
\begin{equation}\label{LDTFP:update t}
	\widehat{t}_{k}=\frac{\left|\boldsymbol{h}_{k}^{\rm H}(\boldsymbol{r})\boldsymbol{S}_{k}\boldsymbol{f}\right|^{2}}{\|\widetilde{\boldsymbol{H}}_{k}^{\rm H}(\boldsymbol{r})\boldsymbol{f}\|_{2}^{2}-\left|\boldsymbol{h}_{k}^{\rm H}(\boldsymbol{r})\boldsymbol{S}_{k}\boldsymbol{f}\right|^{2}+\sigma_{n}^{2}}.
\end{equation}
Note that this solution is consistent with~\eqref{update t:strong duality}, providing an additional validation for strong duality.

\emph{2) Optimization of $\boldsymbol{z}$:} The convex subproblem to optimize $z_{k}$ can be expressed as
\begin{align}
	\underset{z_{k}}{\max}~~&2\sqrt{t_{k}+1}\mathrm{Re}\left\lbrace z_{k}^{\ast}\boldsymbol{h}_{k}^{\rm H}(\boldsymbol{r})\boldsymbol{S}_{k}\boldsymbol{f}\right\rbrace\\
	&-\left|z_{k}\right|^{2}\mathrm{ln}2\left(\|\widetilde{\boldsymbol{H}}_{k}^{\rm H}(\boldsymbol{r})\boldsymbol{f}\|_{2}^{2}+\sigma_{n}^{2}\right).
\end{align}
Leveraging the first-order condition, we can update $z_{k}$ as
\begin{equation}\label{LDTFP:update z_k}
	\widehat{z}_{k}=\frac{\sqrt{t_{k}+1}\boldsymbol{h}_{k}^{\rm H}\boldsymbol{S}_{k}\boldsymbol{f}}{\mathrm{ln}2\left(\|\widetilde{\boldsymbol{H}}_{k}^{\rm H}(\boldsymbol{r})\boldsymbol{f}\|_{2}^{2}+\sigma_{n}^{2}\right)}.
\end{equation}

\emph{3) Optimization of $\boldsymbol{f}$:} By introducing $\nu_{2}$ as the dual variable associated with the constraint $\|\boldsymbol{f}\|_{2}^{2}\leq P_{\rm max}$, we can express the subproblem to optimize $\boldsymbol{f}$ as
\begin{align}
	\underset{\boldsymbol{f}}{\max}~~&{\textstyle\sum}_{k=1}^{K}\big[2\sqrt{t_{k}+1}\mathrm{Re}\left\lbrace z_{k}^{\ast}\boldsymbol{h}_{k}^{\rm H}(\boldsymbol{r})\boldsymbol{S}_{k}\boldsymbol{f}\right\rbrace\nonumber\\
	-&\left|z_{k}\right|^{2}\mathrm{ln}2(\|\widetilde{\boldsymbol{H}}_{k}^{\rm H}(\boldsymbol{r})\boldsymbol{f}\|_{2}^{2}+\sigma_{n}^{2})\big]-\omega\big(\frac{\|\boldsymbol{f}\|_{2}^{2}}{\eta_{\rm PA}}+P_{\rm C}\big)\nonumber\\
	-&\frac{1}{2\mu_{2}}\|\boldsymbol{f}-\widetilde{\boldsymbol{f}}\|_{2}^{2}-\nu_{2}\left(\|\boldsymbol{f}\|_{2}^{2}-P_{\rm max}\right).
\end{align}
Based on the KKT conditions, we can calculate the optimal $\widehat{\boldsymbol{f}}$ as
\begin{equation}\label{LDTFP:update f}
	\widehat{\boldsymbol{f}}=\boldsymbol{\Psi}^{-1}\boldsymbol{q},
\end{equation}
where the auxiliary variables $\boldsymbol{\Psi}$ and $\boldsymbol{q}$ can be expressed as
\begin{equation}\label{LDTFP:calculate Psi}
	\boldsymbol{\Psi}=\sum_{k=1}^{K}\mathrm{ln}2\left|z_{k}\right|^{2}\widetilde{\boldsymbol{H}}_{k}(\boldsymbol{r})\widetilde{\boldsymbol{H}}_{k}^{\rm H}(\boldsymbol{r})+\left(\frac{\omega}{\eta_{\rm PA}}+\frac{1}{2\mu_{2}}+\nu_{2}\right)\boldsymbol{I},
\end{equation}
\begin{equation}\label{LDTFP:calculte q}
	\boldsymbol{q}=\sum_{k=1}^{K}\sqrt{t_{k}+1}z_{k}\boldsymbol{S}_{k}^{\rm H}\boldsymbol{h}_{k}(\boldsymbol{r})+\frac{1}{2\mu_{2}}\widetilde{\boldsymbol{f}}.
\end{equation}
Note that we can obtain $\nu_{2}$ through the bisection method.

The optimization of $\boldsymbol{F}_{\rm RF}$ and $\boldsymbol{F}_{\rm BB}$, as well as the update of $\mu_{2}$, follows the same approach as previously described.

Finally, the LDTFP scheme for the HBF design with fixed RCs can be summarized in \textbf{Algorithm~\ref{Algorithm:LDTFP-HBF}}.

\textbf{Remark 2:} The main difference between the DQTFP and LDTFP schemes lies in the optimization of $\boldsymbol{f}$. 
For the LDTFP scheme, the inner SINR fractional form is extracted from $\mathrm{log}_{2}(1+\widetilde{\gamma}_{k})$, thereby enabling the derivation of the closed-form solution and reducing computational complexity. 
However, the transformation from \eqref{max:EE:PDD} to \eqref{max:EE:DBT+epi} using the DBT operation is not equivalent because of the presence of the penalty term. The decomposition of the subproblems for alternately optimizing $\boldsymbol{t}$ and $\{\boldsymbol{F},\boldsymbol{F}_{\rm RF},\boldsymbol{F}_{\rm BB}\}$ can also lead to some performance loss due to suboptimal results.
In contrast, for the DQTFP scheme, the two sequential QT operations to transform~\eqref{max:EE:PDD} into~\eqref{max:EE:DQTFP} are equivalent, thus enabling optimal solutions.
However, the DQTFP scheme requires solving a convex problem based on the CVX toolbox in each iteration, leading to high computational complexity. 

\textbf{Remark 3:} Both the DQTFP and LDTFP schemes can be readily embedded into the TLAO framework presented in Section~\ref{section:THB SR}, where the HBF design and the RC selection are iteratively optimized. Due to the COD method used in the outer loop, the number of running times scales with $N_{\rm T}$ and $N_{\rm EM}$. As a result, the inner-loop algorithm design requires a trade-off between performance and computational complexity. In Section~\ref{section:simulation}, we will compare the performance and computational complexity of the DQTFP and LDTFP schemes based on simulation results.

\begin{algorithm}[!t]
	\caption{LDTFP Scheme}
	\label{Algorithm:LDTFP-HBF}
	\begin{algorithmic}[1]
		\STATE \textbf{Input:} $N_{\rm T}$, $N_{\rm RF}$, $K$, $\sigma_{n}^{2}$, $P_{\rm max}$, $P_{\rm C}$, and $\eta_{\rm PA}$.
		\STATE Initialize $\boldsymbol{F}$, $\boldsymbol{F}_{\rm RF}$, and $\boldsymbol{F}_{\rm BB}$ via~\eqref{min:WMMSE:initialize F}$\sim$\eqref{min:WMMSE:initialize FBB}.
		\REPEAT
		\REPEAT
		\STATE Update $\omega$ via~\eqref{LDTFP:update omega}.
		\STATE Update $\boldsymbol{t}$ via~\eqref{LDTFP:update t}.
		\STATE Update $\boldsymbol{z}$ via~\eqref{LDTFP:update z_k}.
		\STATE Calculate $\boldsymbol{\Psi}$ and $\boldsymbol{q}$ via~\eqref{LDTFP:calculate Psi} and~\eqref{LDTFP:calculte q}, respectively. Update $\boldsymbol{F}$ via~\eqref{LDTFP:update f}, and obtain $\nu_{2}$ through the bisection method.
		\STATE Update $\boldsymbol{F}_{\rm RF}$ and $\boldsymbol{F}_{\rm BB}$.
		\UNTIL{$\eta_{\rm EE}$ is converged.}
		\STATE Update $\mu_{2}$ as $\mu_{2}\gets\mu_{2}c_{2}$, where we set $0<c_{2}<1$.
		\UNTIL{$\eta_{\rm EE}$ is converged.}
		\STATE \textbf{Output:} $\boldsymbol{F}_{\rm RF}$, $\boldsymbol{F}_{\rm BB}$, and $\eta_{\rm EE}$.
	\end{algorithmic}
\end{algorithm}

\section{Simulation Results}\label{section:simulation}
In this section, we evaluate the performance of the proposed RCRAA-based THBF architecture and optimization schemes. We use the FPA-based architecture and the fully-digital beamformer (FDBF)~\cite{chen2025remaa} as benchmarks. 
Unless otherwise specified, we set the simulation parameters as follows. The numbers of selectable RC candidate points, antenna ports, RF chains and users are set to $N_{\rm EM}=80$, $N_{\rm T}=8$, and $N_{\rm RF}=K=4$, respectively. The spacing between adjacent selectable RC candidate points is set to $d_{\rm p}=\lambda/10$. The channels between the BS and each user are all established with $L=4$ paths. The channel gain of each path obeys $\beta^{(l)}\sim\mathcal{CN}(0,1)$, for $l=1,\cdots,L$. The noise power is set to $\sigma_{n}^{2}=10\mathrm{dBm}$.

\begin{figure}[!t]
	\begin{center}
		\includegraphics[width=80mm]{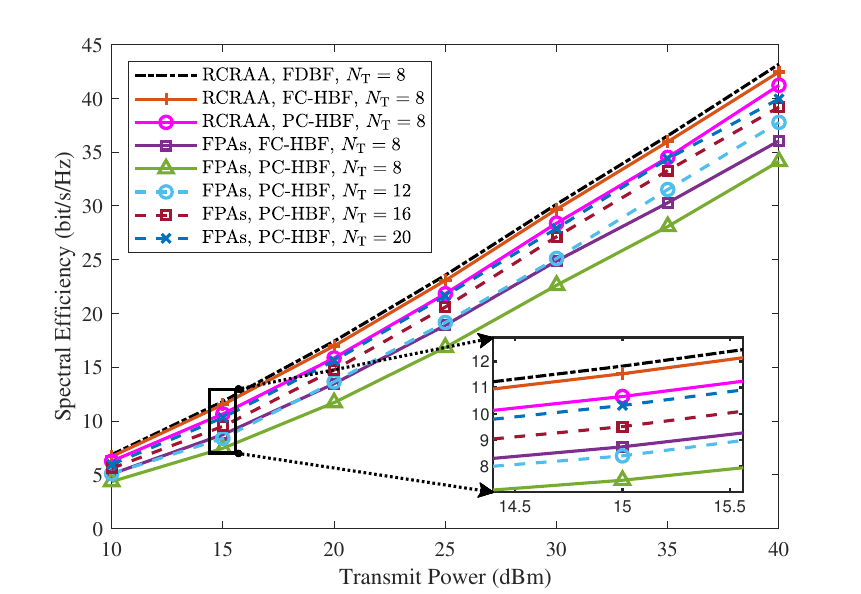}
	\end{center}
	\caption{Comparisons of the SE performance for different architectures with varying transmit powers.}\label{fig:sum-rate PT}
\end{figure}

Fig.~\ref{fig:sum-rate PT} illustrates the SE performance of different architectures for varying transmit powers based on the TLAO-SE scheme, where the transmit power constraint~\eqref{max:sum-rate:PT constraint} is equally satisfied~\cite{zhao2023rethinking}. 
We can observe that the RCRAA architecture achieves better SE performance than the FPA one. The increase in transmit power leads to growing input signal-to-noise ratio (SNR), making the MUI the dominant factor affecting SE performance.
Since the RCRAA offers larger design DoFs than FPAs for mitigating the MUI and adapting better EM environments, the performance gain becomes significant when the transmit power increases.
Besides, the SE of the RCRAA-based FC-HBF and PC-HBF closely approach that of the FDBF benchmark, which validates the effectiveness of the proposed scheme. We can also observe that the performance gap between the FC-HBF and PC-HBF decreases when using the RCRAA. For instance, when the transmit power is $30\mathrm{dBm}$, the performance losses of the PC-HBF compared to the FC-HBF for the RCRAA and FPA architectures are $4.34\%$ and $9.08\%$, respectively. This demonstrates that the RCRAA can effectively compensate for the DoFs lacking in the PC-HBF. 
For further comparisons, we increase the number of antenna ports for the FPA-based architecture to $N_{\rm T}=12$, $N_{\rm T}=16$, and $N_{\rm T}=20$. It is seen that even employing more FPAs at the cost of greatly increased hardware complexity and power consumption, the SE performance is worse than that of the RCRAA with only $N_{\rm T}=8$ antenna ports. This demonstrates the significant advantages of the RCRAA-based THBF architecture over the FPA-based architecture in terms of cost-effectiveness and design flexibility.

\begin{figure}[!t]
	\begin{center}
		\includegraphics[width=80mm]{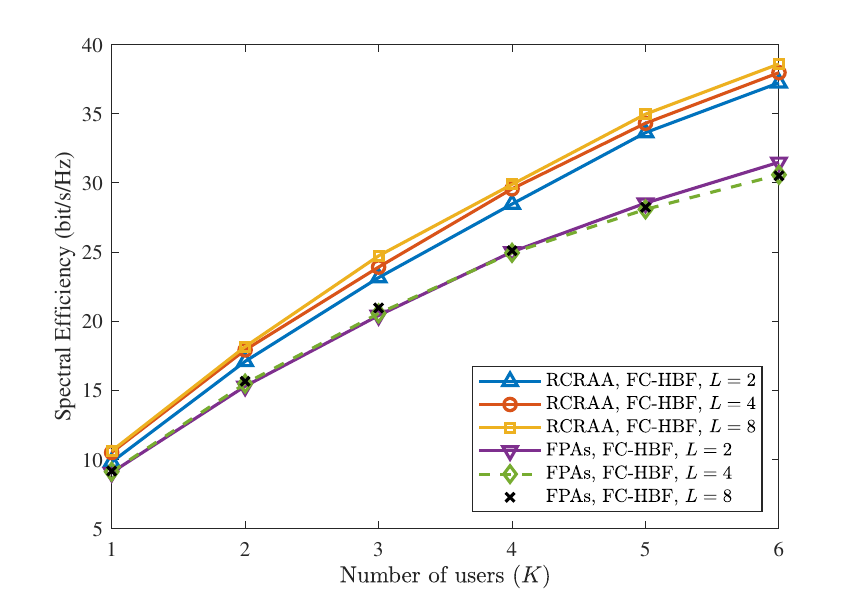}
	\end{center}
	\caption{Comparisons of the SE performance for different architectures with varying numbers of users and channel paths.}\label{fig:sum-rate K}
\end{figure}

Fig.~\ref{fig:sum-rate K} illustrates the SE performance of different architectures for varying numbers of users and channel paths, where we set $P_{\rm max}=30\mathrm{dBm}$. The results show that the performance gap between the RCRAA and FPAs enlarges as the numbers of users and channel paths increases, since the disparity of channel gains becomes significant across the transmit region. As a result, the flexible RC selection in the RCRAA can adapt to complex channel conditions with increased numbers of users and channel paths, thereby improving the SE performance.

\begin{figure}[!t]
	\begin{center}
		\includegraphics[width=80mm]{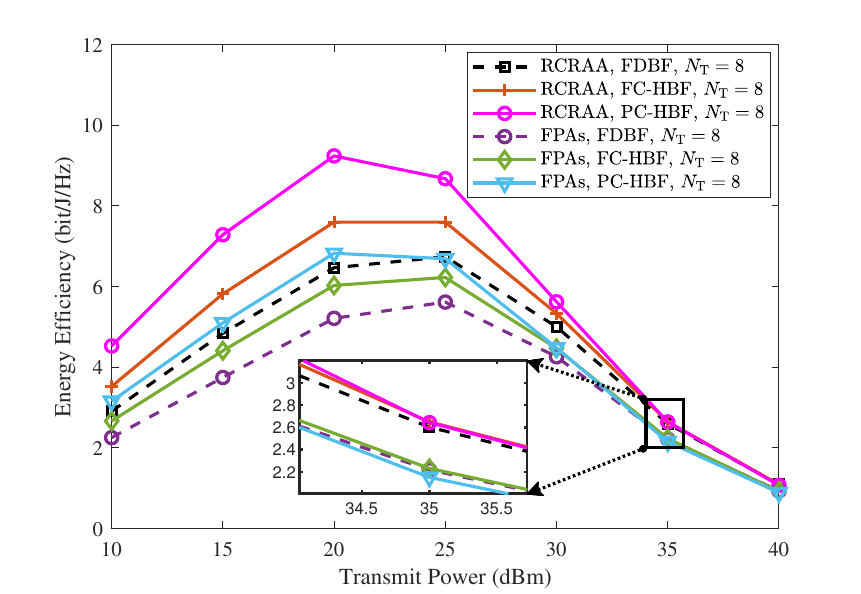}
	\end{center}
	\caption{Comparisons of the EE performance according to the optimized SE for different architectures with varying transmit powers.}\label{fig:EE PT sum-rate max}
\end{figure}

Then, we 
evaluate the EE performance according to the optimized SE, as shown in Fig.~\ref{fig:EE PT sum-rate max}. We set $\eta_{\rm PA}=0.27$, $P_{\rm LO}=22.5\mathrm{mW}$, $P_{\rm RF}=31.6\mathrm{mW}$, $P_{\rm DAC}=128\mathrm{mW}$, and $P_{\rm PS}=21.6\mathrm{mW}$~\cite{Ribeiro2018Energy},~\cite{Mo2017Hybrid}. 
It is seen that all the EE curves first increase and then decrease when enlarging the transmit power, and have the largest value at $P_{\rm T}=20\mathrm{dBm}$ or $25\mathrm{dBm}$. 
If the transmit power is sufficiently large, all the architectures suffer from dramatic EE degradation, since the transmit power dominates in the total power consumption. In particular, the RCRAA architecture achieves better EE performance than the FPA one. Among the FC-HBF, PC-HBF and FDBF, the PC-HBF has the largest EE.

\begin{figure}[!t]
	\begin{center}
		\includegraphics[width=80mm]{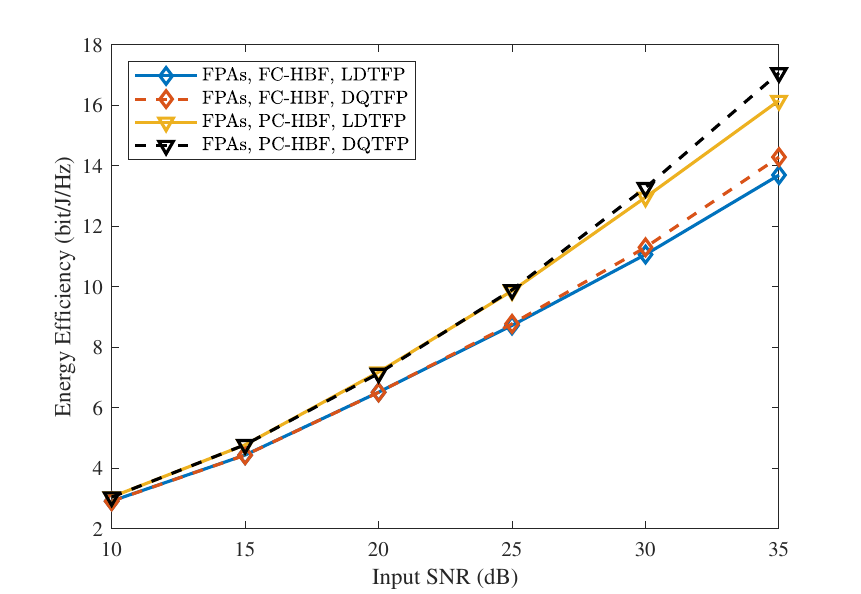}
	\end{center}
	\caption{Comparisons of the LDTFP and DQTFP schemes for EE maximization.}\label{fig:EE_SNR_twomethods}
\end{figure}

Then, we evaluate the proposed EE maximization schemes.
We compare the EE performance achieved by the LDTFP and DQTFP schemes, as shown in Fig.~\ref{fig:EE_SNR_twomethods}. Taking the FC-HBF with 20dB input SNR as an example, i.e., $\sigma_{n}^{2}=10\mathrm{dBm}$, for the configuration of an Intel i7-14650HX CPU, 32GB, and MATLAB R2024a, the execution times of the LDTFP and DQTFP algorithms are $7.68\mathrm{ms}$ and $12.23\mathrm{s}$, respectively. When the input SNR is lower than 25dB, it can be seen that the LDTFP scheme achieves comparable EE performance to the DQTFP scheme, where the former can reduce the execution time by three orders of magnitude of the latter. Therefore, we embed the LDTFP scheme into the TLAO framework to design the THBF for EE maximization.

\begin{figure}[!t]
	\begin{center}
		\includegraphics[width=80mm]{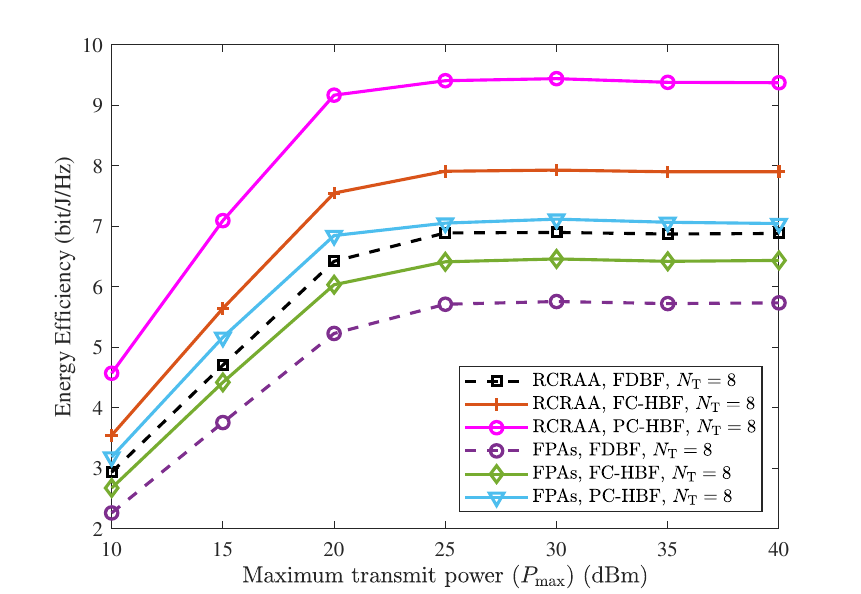}
	\end{center}
	\caption{Comparisons of the EE performance for different architectures with varying maximum transmit powers.}\label{fig:EE_PTmax}
\end{figure}

Fig.~\ref{fig:EE_PTmax} illustrates the EE performance of different architectures for different maximum transmit powers. 
It is seen that the optimal EE performance first increases and then remains constant as the maximum transmit power increases, 
since the transmit power constraint~\eqref{max:EE:PT constraint} becomes loose when $P_{\rm max}$ is sufficiently large. The PC-HBF can achieve the best EE performance due to reduced numbers of RF chains and PSs. Moreover, the EE performance gains of the RCRAA-based FDBF, FC-HBF, and PC-HBF are $19.87\%$, $22.75\%$, and $32.65\%$, respectively, which can validate the significant compensatory effect of the RCRAA on the limited DoFs of the PC-HBF.

\begin{figure}[t]
	\begin{center}
		\includegraphics[width=80mm]{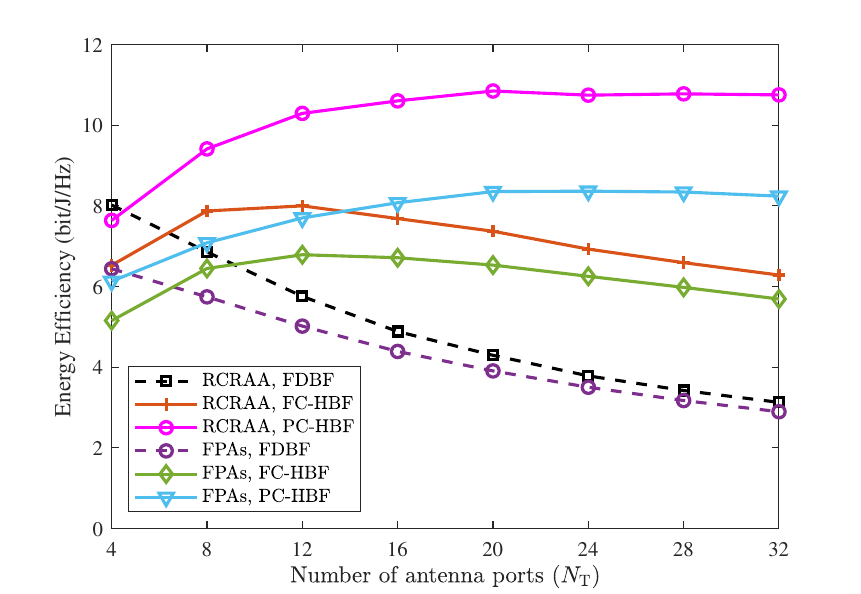}
	\end{center}
	\caption{Comparisons of the EE performance for different architectures with varying number of antenna ports.}\label{fig:EE_NT}
\end{figure}

Fig.~\ref{fig:EE_NT} illustrates the EE performance of different architectures for varying numbers of antenna ports. We can observe that the EEs of both RCRAA and FPA architectures with FDBF and FC-HBF significantly decrease when employing more than $N_{\rm T}=12$ antenna ports. Both architectures with PC-HBF maintain superior EE performance even with large-scale arrays, 
since the PC-HBF greatly reduces the power consumption of RF components.
In addition, for the large-scale array using $N_{\rm T}=32$ antenna ports, the EE performance gains achieved by the RCRAA-based FDBF, FC-HBF and PC-HBF are $8.00\%$, $10.45\%$, and $30.46\%$, respectively. As a result, the RCRAA-based PC-HBF has significant advantage in enhancing the EE performance for large-scale arrays, thereby exhibiting potential for future wireless systems.

\section{Conclusion}\label{section:conclusion}
In this paper, we have proposed the RCRAA-based THBF architecture for multiuser mmWave communications, where the EM beamformer design has been modeled as the RC selection. 
Aiming at SE maximization subject to the hardware and power consumption constraints, we have proposed the TLAO scheme for the THBF design, where the digital and analog beamformers have been optimized based on the PDD in the inner and middle loops, and the RC selection has been determined through the COD method in the outer loop. Aiming at EE maximization, we have developed the DQTFP scheme to tackle the nested fractional form, where the TLAO scheme can be readily used for the THBF design. To reduce the computational complexity, we have proposed the LDTFP scheme, which can significantly reduce the computational complexity with only minor performance loss compared to the DQTFP scheme.
For future research, we will extend our work to different types of reconfigurable antennas.

\bibliographystyle{IEEEtran} 
\bibliography{IEEEabrv,IEEEexample}

\end{document}